\documentclass[journal,twoside,web]{ieeecolor}
\usepackage{generic}
\usepackage{cite}
\usepackage{amsmath,amssymb,amsfonts}
\usepackage{algorithmic}
\usepackage{graphicx}
\usepackage{textcomp}
\def\BibTeX{{\rm B\kern-.05em{\sc i\kern-.025em b}\kern-.08em
    T\kern-.1667em\lower.7ex\hbox{E}\kern-.125emX}}

\usepackage{graphicx, color, xcolor} % for pdf, bitmapped graphics files
\usepackage{epsfig} % for postscript graphics files
\usepackage{amsmath, bbm, bm} 
\usepackage{amsfonts, amssymb}%special font
\usepackage{mathrsfs}
\usepackage{anyfontsize}
\usepackage{soul}

\usepackage{tikz}
\usepackage{pgfplots}
\pgfplotsset{compat=newest}
\usetikzlibrary{plotmarks}
\usetikzlibrary{arrows.meta}
\usepgfplotslibrary{patchplots}
\usepackage{grffile}
\usepackage{subcaption}

\usepackage[hyphens]{url}
\usepackage{hyperref}

\usetikzlibrary{decorations.markings}
\usetikzlibrary{patterns}

\newtheorem{proposition}{Proposition}

\newtheorem{theorem}{Theorem}

\newtheorem{lemma}{Lemma}

\usepackage{tikz}

\usepackage{mathtools}

\let\labelindent\relax

\newtheorem{Definition}{Definition}
\newtheorem{Assumption}{Assumption}
\newtheorem{Example}{Example}
\newtheorem{Counterexample}{Counterexample}
\newtheorem{Remark}{Remark}
\usepackage{enumitem}

\title{\LARGE \bf
 Pairwise Comparison Evolutionary Dynamics with Strategy-Dependent Revision Rates: Stability and $\delta$-Passivity \\ (Expanded Version)}

\author{Semih Kara, \and Nuno C. Martins \thanks{The authors are with the 
Department of Electrical \& Computer 
Engineering and 
the Institute for Systems Research, the
University of Maryland, College Park, MD 20742. 
Email: \{skara@terpmail., nmartins@\}umd.edu. This work is supported in part by 
AFOSR Grant FA9550-19-1-0315.}
}

\begin{document}

  \maketitle
\thispagestyle{empty}
\pagestyle{plain}

%%%%%%%%%%%%%%%%%%%%%%%%%%%%%%%%%%%%%%%%%%%%%%%%%%%%%%%%%%%%%%%%%%%%%%%%%%%%%%%%
\begin{abstract}
We report on new stability conditions for evolutionary dynamics in the context of population games. We adhere to the prevailing framework consisting of many agents, grouped into populations, that interact noncooperatively by selecting strategies with a favorable payoff. Each agent is repeatedly allowed to revise its strategy at a rate referred to as revision rate. Previous stability results considered either that the payoff mechanism was a memoryless potential game, or allowed for dynamics (in the payoff mechanism) at the expense of precluding any explicit dependence of the agents' revision rates on their current strategies. Allowing the dependence of revision rates on strategies is relevant because the agents' strategies at any point in time are generally unequal. To allow for strategy-dependent revision rates and payoff mechanisms that are dynamic (or memoryless games that are not potential), we focus on an evolutionary dynamics class obtained from a straightforward modification of one that stems from the so-called impartial pairwise comparison strategy revision protocol. Revision protocols consistent with the modified class retain from those in the original one the advantage that the agents operate in a fully decentralized manner and with minimal information requirements \textendash they need to access only the payoff values (not the mechanism) of the available strategies. Our main results determine conditions under which  system-theoretic passivity properties are assured, which we leverage for stability analysis. 

\end{abstract}

\section{Introduction}

In this article, we investigate methods to characterize the stability of a continuous-time dynamical system that models the dynamics of noncooperative strategic interactions among the members of large populations of bounded rationality agents. Each agent follows one strategy at a time, but repeatedly (at instants called revision opportunity times) it is allowed to reassess its choice to decide whether to follow a different strategy offering a higher payoff.   The decisions of the agents are coupled by a mechanism that determines the payoff vector, whose entries are the payoffs of the strategies available to the populations. We refer to the rate with which the revision opportunity times occur for an agent as revision rate. In \S\ref{subsec:RevisionRates} we describe the revision rate concept in more detail because it is central to our main results.

\subsection{Overview Of The Technical Framework And Goals}
In our analysis, we adopt the deterministic approach described in~\cite{Park2019From-Population,Park2018Payoff-Dynamic-}, which generalizes that used in most previous work to study population games~\cite{Sandholm2010Population-Game} and evolutionary games~\cite{Weibull1995Evolutionary-ga,Hofbauer2003Evolutionary-ga}. As is explained in~\cite[Section~III]{Park2018Payoff-Dynamic-} and~\cite{Quijano2017The-role-of-pop,Pantoja2011A-population-dy}, the approach is well-suited to analyze large multi-agent systems for which determining the set of stable equilibria (of the dynamical model used) is important because it is a predictor of the long term aggregate strategic behavior of the agents. Specifically, we seek to obtain a systematic methodology to establish {\it global asymptotic stability} {\bf (GAS)} of the said equilibria for a type of payoff mechanism denoted as $\delta$-antipassive~\cite{Fox2013Population-Game}, or more generally $\delta$-antidissipative~\cite{Arcak2020Dissipativity-T}. Important particular cases of these types of payoff mechanism include contractive\footnote{Contractive games were originally called stable games in~\cite{Hofbauer2007Stable-games}. The possibility that calling games stable could cause confusion with notions of system-theoretic stability prompted the nomenclature change.} games or, more generally, weighted contractive~\cite{Arcak2020Dissipativity-T} games and their appropriate dynamic modifications~\cite{Fox2013Population-Game}, which we will later define and call {\it payoff dynamic models} {\bf (PDM)}~\cite{Park2018Payoff-Dynamic-,Park2019From-Population}.

\subsection{Existing Work For The IPC Protocol}
\label{subsec:OverviewImpartial}

Although the above-mentioned work for $\delta$-antipassive and $\delta$-antidissipative PDMs is rather general, it presumes that the agents' revision rates do not depend directly on their current strategies. In the dynamical model, this constraint is present in the bounded rationality rules (or protocols) describing the process by which the agents revise their strategies. The so-called {\it impartial pairwise comparison} {\bf (IPC)}~\cite[\S 7.1]{Hofbauer2009Stable-games-an} protocol, which is particularly relevant for this article\footnote{As we will explain later in Remark~\ref{rem:WhenRM-PCIsIPC}, the IPC class will be a particular case of the protocols analyzed in this article.}, has this limitation. The qualifier {\it impartial} is introduced in~\cite[\S 7.1]{Hofbauer2009Stable-games-an} to indicate that the  revision rates may depend on the current strategy only indirectly through its payoff. Specifically, under an IPC protocol, two agents will have the same revision rate when their strategies have the same payoff. 

%\subsection{Existing Work For Pairwise Comparison Protocols and Potential Games}
%\label{subsec:OverviewPotential}

%At the expense of restricting the payoff mechanism to be a memoryless potential game~\cite{Monderer1996Potential-games} it is possible to use Lyapunov theory~\cite{Sandholm2001Potential-games}\footnote{See also~\cite{Hofbauer2000From-Nash-and-B,Hofbauer1988The-theory-of-e}.} to study the stability of the evolutionary dynamics stemming from a general (not impartial) pairwise comparison protocol~\cite{Sandholm2010Pairwise-compar}. 

%Commenting on these results is important because, as it will become clear in~\S\ref{sec:RMIPC}, the protocol class we will consider to allow for strategy-dependent revision rates is a particular case of the more general pairwise comparison class. 

%\remind{Refer to classes of protocols}
\subsection{Motivation And Objectives}

At the expense of restricting the payoff mechanism to be a memoryless potential game~\cite{Monderer1996Potential-games} it is possible to use Lyapunov theory~\cite{Sandholm2001Potential-games}\footnote{See also~\cite{Hofbauer2000From-Nash-and-B,Hofbauer1988The-theory-of-e}.} to study the stability of the evolutionary dynamics stemming from a general (not necessarily impartial) pairwise comparison protocol~\cite{Sandholm2010Pairwise-compar}. Hence, from~\cite{Sandholm2001Potential-games} and the work discussed in~\S\ref{subsec:OverviewImpartial}, we conclude that existing stability results involving pairwise comparison protocols will either {\bf (i)}~restrict the payoff mechanism to be a memoryless potential game, or {\bf (ii)}~for payoff mechanisms that are dynamic (such as $\delta$-antidissipative PDMs) or memoryless weighted contractive games (which includes concave potential games as a particular case) require the IPC protocol.

In this article, we seek to obtain results that would bridge the gap between (i) and (ii). Specifically, we will generalize the approaches in~\cite{Fox2013Population-Game,Arcak2020Dissipativity-T}, which consider the types of payoff mechanisms mentioned in (ii), so as to allow pairwise comparison protocols that are not necessarily impartial. Our focus on pairwise comparison protocols~\cite{Sandholm2010Pairwise-compar} is justified by their desirable {\it incentive properties}~\cite[\S2.5]{Sandholm2010Pairwise-compar} and inherently fully decentralized operation. As a case in point, the so-called {\it Smith} (pairwise comparison) protocol~\cite{Smith1984The-stability-o} has been widely used to study traffic assignment problems.

\subsection{Preview Of RM-PC Protocols And Main Contributions}
In order to allow for pairwise comparison protocols with strategy-dependent revision rates, in \S\ref{sec:RMIPC} we propose a straightforward modification of the IPC protocol class, which will be referred to throughout this article as the {\it rate-modified pairwise comparison} {\bf (RM-PC)} protocol class. The section also includes a key theorem used in \S\ref{subsec:RM-EPTPCGAS} to specify conditions on the revision rates for which suitable stability properties are assured. Specifically, when the payoff mechanism is a weighted contractive game\footnote{In \S\ref{subsec:WeightedContractivity}, we will revisit and adapt to our context the concepts of weighted contractive game and $\delta$-antidissipative PDM.} or a $\delta$-antidissipative PDM, our technical approach uses system-theoretic passivity~\cite{Fox2013Population-Game} concepts to leverage the results in~\cite{Arcak2020Dissipativity-T} to guarantee for RM-PC protocols satisfying the said conditions that the Nash equilibria set (appropriately defined for the payoff mechanism) is GAS. The {\it hassle~vs~price} game example described in \S\ref{sec:framework}, in which allowing strategy-dependent revision rates will be essential, will illustrate the relevance of our results throughout the article.

\noindent {\bf Informational Requirements:} Protocol classes have inherent informational requirements for implementation~\cite[\S2.3]{Sandholm2010Pairwise-compar}. It will be clear from~\S\ref{sec:RMIPC} that an agent needs only the payoff vector to implement an RM-PC protocol~(see also Remark~\ref{rem:RMPCDecentralized}). Namely, an agent with access to the payoff vector can implement an RM-PC protocol in a decentralized manner without any knowledge about how the payoff vector is generated and it also does not require any information about the strategic choices of the other agents.

\section{Framework Description And Motivation}
\label{sec:framework}

In our framework, each agent belongs to one out of a finite number of populations $\{1,\ldots,\rho\}$, and each agent follows one strategy at a time, which it can change when given a revision opportunity. At every instant, each strategy has a payoff and, at the revision opportunity times, the agents are more likely to switch to strategies whose payoff is higher. Although the set of available strategies is the same for the members of a population, the agents can concurrently follow distinct strategies. 

\subsection{Hassle vs. Price Game (HPG) Example}
A motivating example of application of our framework, which we will be invoking throughout this article to illustrate our contributions, is that of a \underline{"hassle~vs.~price" game {\bf (HPG)}}. In this example, each agent operates a machine that uses a component that must be replaced when it fails. There are several manufacturers that make the component to varying degrees of reliability. Specifically, each component has an exponentially distributed lifetime and its failure rate depends on the manufacturer. The available strategies are the manufacturers, and the payoff of each strategy combines two non-positive terms: (i)~a hassle (disruption) cost that increases with the failure rate and (ii)~the price of the component, which is higher for more reliable manufacturers. The revision opportunity time occurs when the component fails and the agent must decide based on the available information, such as the current payoffs ascribed to the strategies, whether to keep the current strategy~(buy again from the same manufacturer) or follow a different strategy~(decide on another manufacturer to buy from). The agents are partitioned into populations, each uniquely associated to a machine type and/or the undertaking for which the machine is used.

In Example~\ref{ex:Game} (in \S\ref{subsec:payoffmechanism}) we will describe in detail a memoryless payoff mechanism for the HPG, and in Appendix~\ref{app:ExampleSmoothedExample} we will describe a PDM that generalizes Example~\ref{ex:Game}.

\subsection{Population State, Social State and Payoff Vector}
The agents of a population, say population~$r$, are nondescript, hence, their strategy choices at time $t$ can be described by the so-called population state $X^r(t)$ whose entries are proportional to the number of agents selecting the available strategies. In most existing work~\cite{Sandholm2010Population-Game}, the sum of the entries of $X^r$ is a positive constant quantifying the population "mass." Although, to simplify our notation, we consider unit mass populations, our results hold for any population mass after appropriate scaling. Consequently, if $N^r$ is the number of agents in population $r$ then $N^r \text{\small $\times$} X_i^r(t)$ is the number of agents following strategy $i$ at time $t$ in population $r$. The state of the $r$-th population takes values in the following simplex: $$\mathbb{X}^r:=\{x^r \in \mathbb{R}_{\geq 0}^{n^r} \ | \ x^r_1 + \ldots + x^r_{n^r} =1 \},$$ where we use $n^r$ to denote the number of strategies.  The payoffs ascribed at time $t$ to the available strategies of population $r$ are the entries of the payoff vector $P^r(t)$. Namely, $P^r_i(t)$ is the payoff of the $i$-th strategy for population $r$ at time $t$. The so-called social state $X(t)$ at time $t$ is the concatenation of the states of all populations at time $t$, and, similarly, $P(t)$ is the concatenation of the payoff vectors of all populations. Hence, $X(t)$ and $P(t)$ take values in $\mathbb{X} : = \mathbb{X}^1 \times \cdots \times \mathbb{X}^{\rho}$ and $\mathbb{R}^n$, respectively, where $n : = n^1 + \cdots+n^{\rho}$.

A causal payoff mechanism determines $P:=\{ P(t) \ | \ t\geq 0 \}$ in terms of $X:=\{ X(t) \ | \ t \geq 0 \}$. The simplest mechanism is memoryless, acting as $\mathcal{F}: X(t) \mapsto P(t)$, where $\mathcal{F}: \mathbb{X} \rightarrow \mathbb{R}^n$ is a continuously differentiable map referred to as {\it game}. The payoff mechanism may be intrinsic to the problem or it may be influenced by one or more coordinators seeking to steer the social state towards desirable configurations. 

\subsection{Strategy-Dependent Revision Rates: Key Concepts}
\label{subsec:RevisionRates}

\subsubsection{Strategy-Dependent Revision Rates} We assume that, for each $i$ in $\{1,\ldots,n^r\}$, a positive constant $\lambda_i^r$ characterizes the rate at which the agents in population $r$ currently following the $i$-th strategy are allowed to revise their strategy. Specifically, the probability that some agent of population $r$ currently following the $i$-th strategy is allowed to revise its strategy within an infinitesimal time interval of duration $\delta$ is $\delta\text{\small $\times$} \lambda_i^r \text{\small $\times$} N^r \text{\small $\times$} X_i^r(t^*)$, where $t^*$ is in the interval and precedes the revision opportunity time~\cite{Park2019From-Population,Park2018Payoff-Dynamic-}. Moreover, the event that a revision opportunity occurs for a given agent during this period is conditionally independent, given its own current strategy, of the revision opportunity events of all other agents. This independence property holds for our HPG since it is safe to assume that once a new component is installed, the time when it fails depends only on its manufacturer and the agent's population, and not on the choices of the other agents or when the components they currently own fail. 

%Hence, it follows from our description of $X$ that it is a piecewise constant process that jumps when an agent revises its strategy. % I am commenting this because I refer to $X$ being a Markov process later on.

We refer to $\{ \lambda_1^r,\ldots,\lambda_{n^r}^r\}$ as the strategy-dependent revision rates for population $r$ and denote the $n^r$-dimensional vector with its $i$-th index given by $\lambda_i^r$ as $\lambda^r$. 

\subsubsection{Revision Protocols}

Following the standard approach in~\cite[Section~4.1.2]{Sandholm2010Population-Game}, the bounded rationality rule governing how the agents in population $r$ revise their strategies is modeled by a Lipschitz continuous map $\mathcal{T}^r: \mathbb{X}^r \times \mathbf{R}^{n^r} \rightarrow \mathbf{R}_{\geq 0}^{n^r \times n^r}$ referred to as {\it the revision protocol}. When the total number of agents is finite, $ \delta  \text{\small $\times$} N^r \text{\small $\times$} X_i^r(t^*)   \text{\small $\times$} \mathcal{T}_{ij}^r(X^r(t^*),P^r(t^*))$ is the probability that some agent of population $r$ switches from strategy $i$ to $j$, with $i \neq j$, during a time interval of infinitesimally small duration~$\delta$ containing $t^*$ (see~\cite[Section~4.1.2]{Sandholm2010Population-Game} for more details). Specifically, although each agent follows one strategy at a time, the switching strategy may be randomized. 
We interpret $\mathcal{T}$ as having the following structure:
\begin{equation}
\label{eq:ProtocolStructure}
\mathcal{T}_{ij}^r (x^r,p^r) = \lambda_i^r \tau_{ij}^r(x^r,p^r), \quad (x^r,p) \in \mathbb{X}^r \times \mathfrak{P} 
\end{equation} where we invoke the fact explained in~\S\ref{subsec:payoffmechanism} that $P$ takes values in a bounded set $\mathfrak{P} \subseteq \mathbb{R}^n$ and $p^r$, $x^r$ are respectively the sub-vectors of $p$ and $x$ corresponding to a possible payoff and population state for population $r$. Equally important, $\tau_{ij}^r(X^r(t^*),P^r(t^*)) $ would quantify the probability that an agent of population $r$ following the $i$-th strategy will switch at time $t^*$ to strategy $j\neq i$, conditioned on the event that it is allowed to revise its strategy at time $t^*$. Here $\tau$ models probabilistically the bounded rationality decision mechanism of the agents and must satisfy:
\begin{equation}
\label{eq:TauNormalization}
\sum_{j=1, j \neq i}^{n^r} \tau_{ij}^r (x^r,p^r) \leq 1, \quad (x^r,p) \in \mathbb{X}^r \times \mathfrak{P}, \ 1 \leq r \leq \rho
\end{equation}  

%Allowing for distinct revision rates is a substantial improvement over the state of the art requiring $\lambda_i^r=\lambda_j^r$ for the protocol classes that we analyze in \S\ref{sec:RMIPC} and~\S\ref{sec:RMEPT}.

\subsubsection{Deterministic Approximation For Very Large $N^r$}
If a game $\mathcal{F}$ determines $P$ from $X$ as $\mathcal{F}:X(t) \mapsto P(t)$ and each agent revising its strategy at time $t^*$ does so based only on information it has about $X(t^*)$ and $P(t^*)$ then $X$ is a Markov jump process for which the deterministic large-population approximation in~\cite{Kurtz1970Solutions-of-Or} applies. Specifically, as the number of agents $N^r$ of each population $r$ tends to infinity, $X(t)$  and $P(t)$ converge in probability to deterministic limits $x(t)$ and $p(t)$ that we denote as {\it mean social state} and {\it deterministic payoff}, respectively. Naturally, we use $x^r_i(t)$ to denote the proportion of agents in population $r$ following strategy $i$ at time $t$ and $p^r_i(t)$ is the payoff ascribed to the $i$-th strategy in population $r$ at time $t$. 

According to~\cite{Park2019From-Population,Park2018Payoff-Dynamic-}, the deterministic limits are well-defined even when the payoff mechanism is a so-called {\it payoff dynamics model} (PDM) whose definition we will include subsequently. Furthermore, $x(t)$ and $p(t)$ are the solutions of the initial value problem of the so-called {\it mean closed loop model} that we will soon describe in~\S\ref{subsec:EDMPDMDCLM}. 

\subsubsection{Modes Of Convergence and Equilibria}
\label{subsubsec:ModesOfConvergence}

Specifically, according to~\cite[Section~V]{Park2018Payoff-Dynamic-} and~\cite[Section~IV.A]{Park2019From-Population}, it follows from \cite[Theorem~2.11]{Kurtz1970Solutions-of-Or} that, as the numbers of the populations' agents tend to infinity, $X$ and $P$ converge in probability to $x$ and $p$ uniformly over any finite time interval. More importantly, the discussions in~\cite[Section~V]{Park2018Payoff-Dynamic-} and~\cite[Appendix~12.B]{Sandholm2010Population-Game} indicate that the convergence of $X$ towards equilibria, in the limit of large populations, can be established by doing so for~$x$. 

These facts justify our decision to investigate the stability (in the GAS sense) of the equilibria of the mean closed loop model.

\subsection{Mean Closed Loop Model And Its Components}
\label{subsec:EDMPDMDCLM}

%\subsubsection{Evolutionary Dynamics Model (EDM)}
It follows that, for the deterministic approximation~\cite[Section~4.1.2]{Sandholm2010Population-Game}, the rate at which a proportion $x_i^r(t^*)$ of the population $r$ currently following strategy $i$ switches to $j$ at time $t^*$ is $x_i^r(t^*) \text{\small $\times$} \mathcal{T}_{ij}^r(x^r(t^*),p^r(t^*))$. Namely, the following {\it evolutionary dynamics model} {\bf (EDM)} governs the dynamics of~$x$:

\begin{equation}\label{eq:EDM-DEF} \dot {x}^r(t) = \mathcal{V}^r\Big( x^r(t),p^r(t) \Big), \quad t\geq 0, \ 1 \leq r \leq \rho
\end{equation}
where each of the $n^r$ components of $\mathcal{V}^r$, say the $i$-th component, is defined as: 
\begin{multline} \label{eq:EDMfromProtocol} \mathcal{V}_i^r \Big( x^r(t),p^r(t) \Big) := \underbrace{\sum_{j=1, j \neq i}^{n^r} \mathcal{T}_{ji}^r\Big( x^r(t),p^r(t) \Big) x_j^r(t)}_{\text{\footnotesize inflow switching to strategy $i$}} \\  - \underbrace{\sum_{j=1,j \neq i}^{n^r} \mathcal{T}^r_{i j}\Big( x^r(t),p^r(t) \Big) x_i^r(t)}_{\text{\footnotesize outflow switching away from strategy $i$}}
\end{multline}

% Start by making correspondence with revision protocol

% Construction for memoryless follows from Sandholm book, and dynamic follows Park
% Say something to the effect that we call mean social state simply as state and deterministic payoff simply as payoff.

% Say that in many situations when $(X,P)$ is a Markov process it can be approximated by the discrete time model:

\subsubsection{Memoryless Payoff Mechanism}
\label{subsec:payoffmechanism}

In the memoryless case, the payoff mechanism is specified by a continuously differentiable game $\mathcal{F}: x(t) \mapsto p(t)$. Notice that since $\mathbb{X}$ is compact and $\mathcal{F}$ is continuous, $p$ will take values in a bounded set $\mathfrak{P}$.

\begin{Example}\label{ex:Game} The payoff mechanism of our HPG example would be characterized~by:

\begin{equation}
\label{eq:examplepayoff} \mathcal{F}_i^r(x) \underset{HPG}{: =} \underbrace{-\beta^r \lambda_i^r}_{ \begin{matrix}\text{\footnotesize hassle} \\ \text{\footnotesize (replacement) cost} \end{matrix}} - \underbrace{\mathcal{C}_i \big ( \mathcal{D}_i(x) \big)}_{ \text{\footnotesize component price}}, \quad x \in \mathbb{X}
\end{equation} 
where 
\begin{itemize}[leftmargin=0.12 in]
  \itemsep0em
    \item[-] $\{\beta^1,\ldots,\beta^\rho\}$ are positive constants quantifying the costs of replacing a component for the respective population,
    \item[-] $\{1,\ldots,\kappa\}$ is the set of available manufacturers (this is also the strategy set equally available to all\footnote{This means that all populations have the same strategy set and same number of strategies ($n^1=\cdots=n^\rho=\kappa$).} populations),
    \item[-] $\{\lambda_1^r,\ldots,\lambda_{\kappa}^r\}$ are the failure rates of the components for the $r$-th population according to manufacturer, which we assume are ordered as $\lambda_1^r>\ldots>\lambda_{\kappa}^r>0$ (manufacturer $\kappa$ makes the most reliable components),
 %   \comment{Can we have $\{\lambda^1_1,\dots,\lambda^1_\kappa\},\dots,\{\lambda^\rho_1,\dots,\lambda^\rho_\kappa\}$?}
    \item[-] $\mathcal{D}: \mathbb{X} \rightarrow [0,\bar{d}]^\kappa$ gives the (effective) demand from each manufacturer as:
\begin{equation}
\mathcal{D}_i(x) : = \sum_{r=1}^{\rho} \alpha^r x_i^r, \quad 1 \leq i \leq \kappa
\end{equation}
\end{itemize}

Here, $\{\alpha^1,\ldots,\alpha^\rho\}$ are positive constants that quantify the relative weight of each population on the demand. These constants may reflect, for instance, the relative sizes of the populations. Finally, $\mathcal{C}_i:\mathbb{R}_{\geq 0} \rightarrow [c_i,\infty)$ is a continuously differentiable surjective function (of the demand) that quantifies the cost of a component made by the $i$-th manufacturer.
\end{Example}

\begin{Assumption}{\bf (Properties of $\mathcal{C}$ for Example~\ref{ex:Game})} \\
\label{Assumpt:AboutC} We assume that $\mathcal{C}$ has the following properties:
\begin{itemize}
	\item[a)] $\{\mathcal{C}_1,\ldots, \mathcal{C}_{\kappa}\}$ are increasing.
	\item[b)] More reliable components are more expensive, i.e., if ${i> j}$ then $\mathcal{C}_i(d) > \mathcal{C}_j(d)$, for $d$ in $[0,\bar{d}]^\kappa$.
\end{itemize}
\end{Assumption}

As we will explain in~\ref{subsec:WeightedContractivity}, the game~(\ref{eq:examplepayoff}) will satisfy a soon to be defined weighted contractivity property when $\mathcal{C}$ satisfies Assumption~\ref{Assumpt:AboutC}.a. In economic theory, Assumption~\ref{Assumpt:AboutC}.a is referred to as demand-pull inflation~\cite{Perry2017} that occurs when the supply of a product is limited\footnote{Factors restricting supply may include scarcity of raw materials, when manufacturer strategically opts to limit production to keep prices up (as DRAM manufacturers have been doing in the last 3 years), difficulty in ramping up production fast enough to meet demand and sanctions to name a few.}, the manufacturer discounts the price when the demand is weak (and gradually eliminates the discount as demand rises), or when the manufacturer raises the price with increasing demand as a way to increase profits when the product becomes popular. Higher cost (decrease in payoff) for a strategy with higher demand, as measured by the portion of the population following it, is common in many other applications, such as congestion games~\cite{Beckmann1956Studies-in-the-}.

\begin{Remark}{\bf (A labour-market example)} We could model the effect of the contract value on employee turnover in a way that would lead to another example analogous to Example~\ref{ex:Game}. In such an example, a population's agents would be the businesses wishing to hire and retain an employee for a specific job type. Each population would comprise businesses with comparable characteristics from the employees' viewpoint, such as location, structure and size. The strategies available to a population's agents would be the different types of contracts they can offer. In this case, $\mathcal{C}_i$ in (\ref{eq:examplepayoff}) would determine the cost of contract $i$ as a function of the demand. Cheaper contracts offering worse benefits and/or lower salary would lead to a higher turnover rate (quantified by $\lambda_i^r$) and associated increased cost for retraining and rehiring (quantified by $\beta_i^r \lambda_i^r$).
\end{Remark}
%\remind{Talk about some of the energy-related games ?}

\subsubsection{Payoff Dynamics Model (PDM)}
More generally, the payoff mechanism is modeled by a payoff dynamics model~(PDM) with the following structure:
\begin{equation}
\label{eq:PDM}
\quad \begin{matrix}
\dot{q}(t) = & \mathcal{G} \big (q(t),x(t) \big ) \\
p(t)  = &\mathcal{H} \big (q(t),x(t) \big )
\end{matrix},  \quad t \geq 0, \quad q(0) \in \mathfrak{Q}_0
\end{equation}
where $\mathfrak{Q}_0 \subseteq \mathbb{R}^m$ is a bounded set, $\mathcal{G}: \mathbb{R}^m \times \mathbb{X} \rightarrow \mathbb{R}^n $ is Lipschitz continuous and ${\mathcal{H}: \mathbb{R}^m \times \mathbb{X} \rightarrow \mathbb{R}^n}$ is continuously differentiable and Lipschitz continuous.

Hence, (\ref{eq:PDM}) specifies a PDM that operates as a causal nonlinear dynamical system with input $x$ and output $p$. As discussed in~\cite{Park2019From-Population,Park2018Payoff-Dynamic-}, PDMs can be used to account for dynamic behaviors inherent to certain payoff mechanisms, such as delays, pricing inertia, agent-level learning, and also to isolate long-term trends~\cite{Fox2013Population-Game}. 

The analysis in this article presumes, as was the case in~\cite{Park2019From-Population,Park2018Payoff-Dynamic-}, that the state $q$ remains in a bounded set $\mathfrak{Q}$. Notice that input to state stability~\cite{Sontag1998Mathematical-Co} of the PDM would suffice to guarantee that $q$ remains in a bounded set because $x$ takes values in a bounded set. Furthermore, the fact that $\mathcal{H}$ is Lipschitz continuous also guarantees that $p$ remains in a bounded set $\mathfrak{P}$. Finally, we consider that there is a game $\mathcal{F}_{\mathcal{G},\mathcal{H}}$ that equals $\mathcal{H}$ in the stationary regime:
\begin{equation}
\label{eq:stationarygame}
\mathcal{G}(x,q) = 0 \implies \mathcal{F}_{\mathcal{G},\mathcal{H}}(x)=\mathcal{H}(x,q), \quad  x \in \mathbb{X}, \ q \in \mathbb{R}^n
\end{equation}

In Appendix~\ref{app:ExampleSmoothedExample}, we describe a PDM example constructed as a dynamic modification of Example~\ref{ex:Game}.

\subsubsection{Mean Closed Loop Model}

\tikzstyle{system} = [draw, ultra thick, minimum width=6em, text centered, rounded corners, fill=black!4,  minimum height=3em]

\def\edgedist{1.1}
\begin{figure} [t]
\centering
\begin{tikzpicture}[scale=0.5, node distance=7em,>=latex]

\node [system, minimum height=8ex] (games) {$\begin{array} {c} \textit{Payoff Mechanism} \\ \textit{(PDM) or $\mathcal{F}$} \end{array}$};
\draw [->] (games)+(0,2) node[above,color=green!15!black]{$q(0)$} -- (games);

\node [system, minimum height=8ex, below of=games] (evo_dynamics) {$\begin{array} {c} \textit{Evolutionary Dynamics Model} \\ \textit{(EDM)} \end{array}$};
\draw [->] (evo_dynamics)+(0,-2) node[below,color=green!15!black]{$x(0)$} -- (evo_dynamics);

\draw[ultra thick, ->] (games.east) -- ++(2.8*\edgedist,0) node [pos=.5, above] {\large $p$} |- (evo_dynamics.east) node [above, pos=.25, rotate=-90] {\textit{deterministic payoff}};
\draw[ultra thick,->] (evo_dynamics.west) -- ++(-1*\edgedist,0) node [pos=.5,below] {\large $x$} |- (games.west) node [above, pos=.25, rotate=90] {\textit{mean social state}} ;

\end{tikzpicture}
\caption{Diagram representing a feedback interconnection between a PDM (or game $\mathcal{F}$) and an EDM. The resulting system is referred to as {mean closed loop model}.}
\label{Fig:ClosedLoop}
\end{figure}
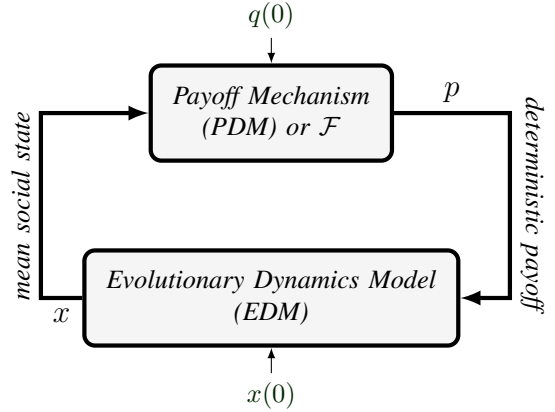

The EDM and the payoff mechanism interact in feedback according to the so-called {\it mean closed loop model} (see~Fig.\ref{Fig:ClosedLoop}), which after substituting~$p(t) = \mathcal{F}(x(t))$ into~(\ref{eq:EDM-DEF}) has the following structure when the payoff mechanism is memoryless and specified by a game~$\mathcal{F}$:

\begin{equation}
\label{eq:closedloopmodelF}
\dot{x}^r(t)  = \mathcal{V}^r \Big( x^r(t), \underbrace{\mathcal{F}^r \big (x(t) \big )}_{p^r(t)} \Big),  \quad r \in \{1,\ldots,\rho\}
\end{equation}

More generally, if the payoff mechanism is a PDM, then after substituting~(\ref{eq:PDM}) into~(\ref{eq:EDM-DEF}) the mean closed loop model is specified as follows:

\begin{subequations}
\label{eq:closedloopmodelPDM}
\begin{align}
\dot{q}(t) &=  \mathcal{G} \big (q(t),x(t) \big ), & t \geq 0\\
\dot{x}^r(t) & = \mathcal{V}^r \Big( x^r(t), \underbrace{\mathcal{H}^r \big (q(t),x(t) \big )}_{p^r(t)} \Big),  & r \in \{1,\ldots,\rho\}
\end{align}
\end{subequations}

%Explain rates, e

\section{Nash Stationarity, Weighted Contractivity, and Problem Formulation}

Our analysis will focus on establishing the global asymptotic stability (GAS) of the equilibria of~(\ref{eq:closedloopmodelF}) or~(\ref{eq:closedloopmodelPDM}) by analysing the solutions of the initial value problem that are guaranteed by the Picard-Lindel\"{o}f theorem to exist and be unique for each $x(0)$ in $\mathbf{X}$, or each pair $\big( x(0), q(0) \big)$ in $\mathbf{X} \times \mathfrak{Q}_0$, respectively.

\subsection{Nash Equilibria Set and Nash Stationarity}
\label{subsec:NE-Intro}

We start by defining the Nash equilibria set for a game $\mathcal{F}$ as follows:
$$ \mathbb{NE}(\mathcal{F}) : = \Big \{ \ x \in \mathbb{X} \ \ \Big | \ \ x^T \mathcal{F}(x) \geq y^T \mathcal{F}(x), \ \ y \in \mathbb{X} \Big \} $$

As explained in~\cite{Sandholm2010Population-Game}, there are important classes of protocols satisfying the so-called Nash stationarity property defined below. Fortunately, as we observe in~\S\ref{sec:RMIPC}, RM-PC protocols are Nash stationary.

\begin{Definition} Given $r$ in $\{1,\ldots,\rho\}$, a protocol for population $r$ satisfies the Nash stationarity property, if the following equivalence holds for the EDM~(\ref{eq:EDMfromProtocol}) for all $p^r$ in $\mathbb{R}^{n^r}$:
\begin{equation} \label{eq:NS} (x^r)^T p^r = \max_{y \in \mathbb{X}^r} y^T p^r \Leftrightarrow \mathcal{V}^r(x^r,p^r) = 0
\end{equation} Thus, Nash stationarity implies that $x^r$ at an equilibrium must be a best response to $p^r$.
\end{Definition}

Hence, if Nash stationarity holds for all populations, then $x$ at an equilibrium of the mean closed loop will be either in $\mathbb{NE}(\mathcal{F})$ when the payoff mechanism is $\mathcal{F}:x(t) \mapsto p(t)$, or $x$ will be in $\mathbb{NE}(\mathcal{F}_{\mathcal{G},\mathcal{H}})$ when the payoff mechanism is a PDM. In these cases, $x$ is guaranteed to converge to a Nash equilibrium when the equilibria set of the mean closed loop is globally asymptotically stable (GAS). Notably, when it is GAS, the Nash equilibria set predicts the long-term behavior of both $x$ and $X$ in the limit of large populations, as noted in~\S\ref{subsubsec:ModesOfConvergence}.

Subsequently, we discuss why GAS assuages some of the well-known criticism of the Nash equilibrium concept and gives it a well-motivated role in our context.

\subsection{Global Asymptotic Stability and Nash Equilibria}
\label{app:GASandNE}

We start by observing that Nash equilibria\footnote{See~\cite{Holt2004The-Nash-equili} for various interpretations of the Nash equilibrium concept.} in our context should be interpreted in the mass-action sense described in~\cite{Weibull1995The-mass-action}, which was originally proposed by Nash in~\cite{Jr.1951Non-Cooperative}.

We proceed by arguing that our results establishing GAS of the Nash equilibria set for our framework may mitigate some of the criticism~\cite{Kreps1990Game-Theory-and} of the Nash equilibrium concept. Specifically, RM-PC protocols governing/modeling the agents' decisions follow bounded rationality rules that rely solely on knowledge of the payoff vector (see~\S\ref{sec:RMIPC} for the informational requirements of RM-PC protocols). 
Hence, notwithstanding the exiguous informational requirements of RM-PC protocols, when the conditions for our GAS results are met, they will assure convergence of $x$ to the Nash equilibria set, in which case the prevalent criticism that Nash equilibria are viable only when the agents know each others' strategies does not apply.

Lack of uniqueness is another common reason to claim that any prediction of the long-term behavior of $x$ based on the Nash equilibria set is uncertain. However, in applications it often suffices to predict that $x$ will satisfy a property shared by all such equilibria. One example is when $\mathcal{F}$ has a concave potential that we seek to maximize, in which case the Nash equilibria are exactly the optima.  Moreover, {\it price of anarchy}~\cite{Roughgarden2005Selfish-routing,Nisam2007Algorithmic-Gam} upper-bounds provide provable guarantees on the degree to which the Nash equilibria are sub-optimal with respect to the population average payoff (see also~\cite[\S3.1.6 and \S3.1.7]{Sandholm2010Population-Game}). Alternatively, if $\mathcal{F}$ is to be used by a coordinator to spur desirable behavior by the population then it may be possible to design it in a way that limits the "size" of the Nash equilibria set.

\subsection{Key Assumptions}
\label{subsec:WeightedContractivity}

In \S\ref{subsec:RM-EPTPCGAS}, we will be able to use the results in \S\ref{sec:RMIPC} in conjunction with~\cite[Corollary~1]{Arcak2020Dissipativity-T} to guarantee the stability of $\mathbb{NE}(\mathcal{F})$ for~(\ref{eq:closedloopmodelF}) under the following assumption.
\begin{Assumption}
\label{ass:WeightedContractiveF} If the payoff mechanism is a memoryless map $\mathcal{F}:x(t) \mapsto p(t)$, then we assume that there are positive weights $\{w^1,\ldots,w^\rho\}$ for which the following holds:
\begin{equation}
\label{eq:weightedcontractive}
\sum_{r=1}^\rho w^r\Big ( \mathcal{F}^r(x^r) -\mathcal{F}^r(\tilde{x}^r) \Big)^T(x^r - \tilde{x}^r) \leq 0, \quad x,\tilde{x} \in \mathbb{X}
\end{equation} The inequality in~(\ref{eq:weightedcontractive}) coincides with contractivity~\cite{Hofbauer2007Stable-games} when the weights are identical, and can be viewed, more generally, as weighted contractivity~\cite{Arcak2020Dissipativity-T} with respect to a block-diagonal matrix $\mathbf{W}:=\mathbf{diag} \ (w^1 \mathbf{I}_{n^1\times n^1},\ldots,w^\rho \mathbf{I}_{n^\rho\times n^\rho})$ with unequal weights.
\end{Assumption}

\begin{Remark} By following an approach analogous to that of~\cite[\S IV.A]{Arcak2020Dissipativity-T}, one can show that Example~\ref{ex:Game} is weighted contractive with $w^r=\alpha^r$, for $r$ in $\{1,\ldots,\rho\}$.
\end{Remark}

More generally, we will be able to use~\cite[Theorem~2]{Arcak2020Dissipativity-T} to ascertain GAS of $\mathbb{NE}(\mathcal{F}_{\mathcal{G,H}})$ for~(\ref{eq:closedloopmodelPDM}) when the payoff mechanism is a PDM~(\ref{eq:PDM}) satisfying the following assumption.

\begin{Assumption} \label{ass:PDMDissipative} If the payoff mechanism is a PDM, then we assume that there are positive weights $\{w^1,\ldots,w^\rho\}$ for which it satisfies the $\delta$-antidissipativity conditions in~\cite[(39)-(40)]{Arcak2020Dissipativity-T} with respect to $\Pi$ constructed as in~\cite[(18)]{Arcak2020Dissipativity-T}.
\end{Assumption}

\begin{Remark} \label{rem:ThePDMExampleWorks}
One can appropriately modify the steps in the proof of~\cite[Proposition~3]{Arcak2020Dissipativity-T} to show that the PDM example described in Appendix~\ref{app:ExampleSmoothedExample} satisfies Assumption~\ref{ass:PDMDissipative} with ${w^r=\alpha^r}$, for $r$ in $\{1,\ldots,\rho\}$.
\end{Remark}

Several additional examples of contractive games, weighted contractive games, $\delta$-antipassive and $\delta$-antidissipative PDMs can be found in~\cite{Sandholm2010Population-Game}, \cite{Arcak2020Dissipativity-T}, [\citenum{Fox2013Population-Game},\citenum{Park2019From-Population},\citenum{Park2018Payoff-Dynamic-}], and~\cite{Arcak2020Dissipativity-T}, respectively.

\subsection{Technical Approach}
In order to leverage the results in~\cite{Arcak2020Dissipativity-T} to establish GAS of $\mathbb{NE}(\mathcal{F})$ for~(\ref{eq:closedloopmodelF}), or GAS of $\mathbb{NE}(\mathcal{F}_{\mathcal{G,H}})$ for~(\ref{eq:closedloopmodelPDM}) we will also need that the protocol for each population is $\delta$-passive according to the following definition.

\begin{Definition}{\bf (protocol $\delta$-passivity)}
\label{def:PWiseDPass}
Given $r$ in $\{1,\ldots,\rho\}$, the  protocol for population $r$ is \underline{$\delta$-passive} if there  are functions $\mathcal{S}^r:\mathbb{X}^r \times \mathbb{R}^{n^r} \rightarrow \mathbb{R}_{\geq 0}$ and $\mathfrak{S}^r:\mathbb{X}^r \times \mathbb{R}^{n^r} \rightarrow \mathbb{R}_{\geq 0}$ such that the following holds:
\begin{subequations}
\label{eq:PWiseDPass}
\begin{multline}
\label{eq:PWiseDPass_ineq}
\tfrac{\partial \mathcal{S}^r(x^r,p^r)}{\partial x^r} \mathcal{V}^r(x^r,p^r)+\tfrac{\partial \mathcal{S}^r(x^r,p^r)}{\partial p^r} u^r  \\ \leq - \mathfrak{S}^r(x^r,p^r)+\mathcal{V}^r(x^r,p^r)^Tu^r
\end{multline}
\begin{align}
\mathcal{S}^r(x^r,p^r) = 0 & \Leftrightarrow \mathcal{V}^r(x^r,p^r) = 0 \label{eq:PWiseDPass_Stat1}\\
\mathfrak{S}^r(x^r,p^r) = 0 & \Leftrightarrow \mathcal{V}^r(x^r,p^r) = 0 \label{eq:PWiseDPass_Stat2}
\end{align}
\end{subequations}
for all $x^r$ in $\mathbb{X}^r$, $p^r,u^r$ in $\mathbb{R}^{n^r}$. Following the convention in~\cite{Park2019From-Population,Park2018Payoff-Dynamic-}, we will refer to $\mathcal{S}^r$ as a \underline{$\delta$-storage function}. Note that $\partial\mathcal{S}^r/\partial x^r$ and $\partial\mathcal{S}^r/\partial p^r$ denote respectively the transpose of the gradient of $\mathcal{S}^r$ with respect to its first and second argument.
\end{Definition}

See~\cite[Remark~3]{Arcak2020Dissipativity-T} for a comparison between $\delta$-passivity as defined above, $\delta$-dissipativity and $\delta$-passivity as proposed in~\cite{Fox2013Population-Game}. One can readily repurpose the proofs of \cite[Theorem~4.5]{Fox2013Population-Game}\footnote{ See \cite[Proposition~4]{Park2018Payoff-Dynamic-} for a more general proof, and~\cite{Park2018Passivity-and-e,Park2018Passivity-and-E2,Park2018Payoff-Dynamic-} for a complete analysis of $\delta$-passivity for this and other protocols.} to conclude that the IPC protocols are $\delta$-passive according to Definition~\ref{def:PWiseDPass}. These conclusions can also be recovered as a particular case of our analysis establishing $\delta$-passivity for the broader class of RM-PC protocols proposed and analyzed in~\S\ref{sec:RMIPC}.

\subsection{Problem Formulation}
\label{subsec:Problemformulation}

We start by defining the following worst-case ratios that will be used throughout this article to quantify the relative discrepancies among the revision rates of each population.

\begin{Definition}
Given $r$ in $\{1,\ldots,\rho\}$ and the revision rates $\{ \lambda_i^r \ | \ 1 \leq i \leq n^r\}$ for population $r$, we define the worst-case revision rate ratio for the $r$-th population as follows:
\begin{equation} \label{eq:WorstRatioDefinition}
    \lambda_R^r : = \max \Bigg \{ \frac{\lambda^r_i}{\lambda^r_j} \ \Bigg | \  i,j \in \{1, \ldots,  n^r\} \Bigg \} % \quad 1 \leq r \leq \rho
\end{equation}
\vspace{.1in}
\end{Definition}

Notice that $\lambda_R^r \geq 1$ holds by definition and $\lambda_R^r = 1$ if and only if the revision rates for the $r$-th population are identical.

In order to develop a methodology that can cope with the case in which $\lambda_R^r > 1$ for one or more populations (unequal revision rates), in~\S\ref{sec:RMIPC} we seek to solve the following subproblems:
\begin{itemize}
\item[i)] Propose practicable modified protocols that are compatible with any pre-selected revision rates. (As we already mentioned, the modified class of protocols RM-PC will be our answer to this subproblem.) 
\item[ii)] Determine conditions on $\{ \lambda_R^r \ | \ 1\leq r \leq \rho\}$, and other parameters, under which the RM-PC protocols are $\delta$-passive. Under the assumptions in~\S\ref{subsec:WeightedContractivity}, this will allow us to leverage~\cite[Corollary~1]{Arcak2020Dissipativity-T} or~\cite[Theorem~2]{Arcak2020Dissipativity-T} to establish GAS of $\mathbb{NE}(\mathcal{F})$ for~(\ref{eq:closedloopmodelF}) or GAS of $\mathbb{NE}(\mathcal{F}_{\mathcal{G,H}})$ for~(\ref{eq:closedloopmodelPDM}), respectively.
\end{itemize}

\section{RM-PC Protocol And Main Results}
\label{sec:RMIPC}

In this section, we address the problem formulation goals listed in~\S\ref{subsec:Problemformulation} for the protocol class we propose below:
\begin{Definition}\label{def:RM-PCprot}{\bf (RM-PC protocol)} Given $r$ in $\{1,\ldots,\rho\}$, the protocol~(\ref{eq:ProtocolStructure}) of the $r$-th population is of the {\it rate-modified} pairwise comparison (RM-PC) class if $\tau^r$ can be written as:
\begin{equation}
\label{eq:tauRMPC}
\tau^r_{ij}(x^r,p^r) = \tfrac{1}{\bar{\tau}^r} \phi^r_{j} (p^r_j -p^r_i)
\end{equation} where $\bar{\tau}^r$ is a positive normalization constant for which~(\ref{eq:TauNormalization}) holds, while $\phi^r_{j}:\mathbb{R} \rightarrow \mathbb{R}_{\geq 0}$ is Lipschitz continuous and sign-preserving, meaning that $\phi^r_{j} (\delta) >0$ for $\delta >0$ and $\phi^r_{j} (\delta)=0$ for $\delta \leq 0$.
\end{Definition}

By substituting~(\ref{eq:tauRMPC}) into (\ref{eq:EDMfromProtocol}), we obtain the following RM-PC EDM model for the $r$-th population for each $i$ in~$ \{1,\dots,n^r\}$:
\begin{align} \nonumber
(\mathcal{V}_i^{\text{\tiny{RM-PC}}})^r (x^r,p^r) := & \sum_{j=1, j \neq i}^{n^r} \lambda_j^r \tfrac{1}{\bar{\tau}^r} \phi^r_{i} (p^r_i -p^r_j) x_j^r \\ \label{eq:EDMDef}
& - \sum_{j=1,j \neq i}^{n^r} \lambda_i^r \tfrac{1}{\bar{\tau}^r} \phi^r_{j} (p^r_j -p^r_i) x_i^r
\end{align}

\begin{Remark} {\bf (IPC is an RM-PC subclass)} \label{rem:WhenRM-PCIsIPC} 
In the particular case in which the revision rates for the $r$-th population are equal ($\lambda_1^r=\cdots=\lambda_{n^r}^r$), an RM-PC protocol becomes of the IPC class considered in previous work characterizing $\delta$-passivity~\cite{Fox2013Population-Game}.
\end{Remark}

% \begin{figure}[h!]
% \centering
% \begin{tikzpicture}[scale=1]
% \definecolor{antiquewhite}{rgb}{0.97, 0.91, 0.81}
% \draw[fill=antiquewhite, fill opacity=0.4, rounded corners] (0,0) rectangle (8.5, 3){};
% \node [at = {(7.35,2.4)}] {\small \begin{tabular}{r} PC: \\ $\frac{1}{\bar{\tau}^r}\phi^r_{ij}(x^r,p^r)$ \end{tabular}};
% \draw[fill=antiquewhite, fill opacity=0.4, rounded corners] (0.3,0.3) rectangle (5.8, 2.5){};
% \node [at = {(4.55, 2)}] {\small \begin{tabular}{r} RM-PC: \\ $\frac{1}{\bar{\tau}^r}\lambda_i^r\phi^r_{j}(p_j^r-p_i^r)$ \end{tabular}};
% \draw[fill=antiquewhite, fill opacity=0.4, rounded corners] (0.6,0.6) rectangle (3, 2){};
% \node [at = {(1.9, 1.5)}] {\small \begin{tabular}{r} IPC: \\ $\frac{1}{\bar{\tau}^r}\phi^r_{j}(p_j^r-p_i^r)$ \end{tabular}};
% \end{tikzpicture}
% \caption{A diagram depicting the relationship between the PC, RM-PC and IPC protocols.}
% \label{fig:PC_inclusion_diagram}
% \end{figure}

\begin{Example}{\bf (RM-Smith protocol)} \label{ex:RM-Smith} As an example of a RM-PC protocol, we can define the {\it rate-modified} Smith protocol (RM-Smith) by substituting $\phi^r_j (\cdot) = [\cdot]_+$ in~(\ref{eq:tauRMPC}) and~(\ref{eq:ProtocolStructure}), leading to: 
\begin{equation}
\label{eq:RM-SmithDef}
\mathcal{T}_{ij}^r (x^r,p^r) \underset{\text{\tiny RM-Smith}}{=} \lambda_i^r \tfrac{1}{\bar{\tau}^r} [p^r_j -p^r_i]_+, \quad (x^r,p) \in \mathbb{X}^r \times \mathfrak{P} 
\end{equation} 
and the following EDM after substitution in~(\ref{eq:EDMDef}):
\begin{align} \nonumber
(\mathcal{V}_i^{\text{\tiny{RM-Smith}}})^r (x^r,p^r) := & \sum_{j=1, j \neq i}^{n^r} \lambda_j^r \tfrac{1}{\bar{\tau}^r} [p^r_i -p^r_j]_+ x_j^r \\ \label{eq:SmithEDMDef}
& - \sum_{j=1,j \neq i}^{n^r} \lambda_i^r \tfrac{1}{\bar{\tau}^r} [p^r_j -p^r_i]_+ x_i^r
\end{align}
Consequently, the probability that, at a revision opportunity time, an agent following the RM-Smith protocol switches from strategy $i$ to $j$ is proportional to the positive part of the payoff difference. Notice that when the revision rates are equal {(${\lambda_1^r=\cdots=\lambda_{n^r}^r}$)} the RM-Smith protocol reduces to the well-known Smith protocol originally proposed in~\cite{Smith1984The-stability-o} to analyze the dynamics of traffic assignment strategies. 
\end{Example}

\begin{Remark}{\bf (RM-PC: Informational Requirements)} 
\label{rem:RMPCDecentralized} It follows from~(\ref{eq:tauRMPC}) that, other than knowledge of the payoff of the available strategies for the population it is a part of, each agent following an RM-PC protocol does not need to coordinate with other agents and it does not require any additional information about the social state or the strategic choices of the other agents.
\end{Remark}

\subsection{RM-PC Protocol: Nash Stationarity And $\delta$-passivity}
\label{subsec:RM-PCNashStatand}

In this subsection, we establish Nash stationarity and identify $\delta$-passivity properties of RM-PC protocols. Theorem~\ref{thm:RM-PC_diss} is the main result of this section, which will allow us to invoke results in~\cite{Arcak2020Dissipativity-T} to draw important conclusions on the stability of the mean closed loop (see~\S\ref{subsec:RM-EPTPCGAS} for more details).

%We commence by introducing the following lemma which identifies Nash stationarity properties of RM-PC protocols.

\subsubsection{Pairwise Comparison Protocols and Nash Stationarity}

%According to \cite[\S~4.1]{Sandholm2010Pairwise-compar}, a revision protocol $\mathcal{T}^r$ is said to be of the pairwise comparison (PC) class if for all $x^r\in\mathbb{X}^{r}$, $p^r\in\mathbb{R}^{n^r}$ and $i,j\in\{1,\dots,n^r\}$ we can write $\mathcal{T}_{ij}^r(x^r,p^r)=\phi^r_{ij}(x^r,p^r)/\bar{\tau}^r$ for some positive normalizing constant $\bar{\tau}^r$ and a Lipschitz continuous function $\phi^r_{ij}:\mathbb{X}^r\times\mathbb{R}^{n^r}\to \mathbb{R}_{\geq 0}$ satisfying $\phi^r_{ij}(x^r,p^r)>0$ if and only if $p^r_j-p^r_i>0$. 

%From this definition, and equations (\ref{eq:TauNormalization}) and (\ref{eq:tauRMPC}) it follows that RM-PC protocols are specific types of PC protocols for which $\phi^r_{ij}(x^r,p^r)=\lambda^r_i\phi^r_{j}(p^r_j-p^r_i)$. Moreover, in the particular case in which the revision rates for the $r$-th population are equal ($\lambda_1^r=\cdots=\lambda_{n^r}^r$), an RM-PC protocol becomes of the IPC class considered in previous work (see~\cite[(18)]{Hofbauer2009Stable-games-an} and~\cite[(65)\textendash(67)]{Fox2013Population-Game}). A diagram summarizing the relationship between protocols of the PC, RM-PC and IPC classes is depicted in Fig.~\ref{fig:PC_inclusion_diagram}.  (2) Although previous contractivity and $\delta$-passivity results presumed IPC protocols, there is existing work establishing other useful properties for the broader PC protocol class. Notably, \cite[Theorem~1]{Sandholm2010Pairwise-compar} states that a PC protocol is Nash stationary, which directly leads to the following corollary.

The RM-PC class is a particular case of the so-called {\it pairwise comparison} protocol class defined in~\cite[\S~4.1]{Sandholm2010Pairwise-compar}. It is relevant to recognize this because, although previous contractivity~\cite{Hofbauer2009Stable-games-an} and $\delta$-passivity~\cite{Fox2013Population-Game} results that we seek to generalize were restricted to IPC protocols only, there is existing work establishing other useful properties for the much broader pairwise comparison protocol class. Notably, \cite[Theorem~1]{Sandholm2010Pairwise-compar} states that a pairwise comparison protocol is Nash stationary, which leads directly to the following lemma.

\begin{lemma} {\bf (RM-PC protocol is Nash stationary)}
\label{lem:RM-PC_NS}
Given $r$ in $\{1,\ldots,\rho\}$, if the $r$-th population's protocol is of the RM-PC class, then~(\ref{eq:NS}) holds for any positive revision rates $\{ \lambda_i^r \ | \ 1 \leq i \leq n^r\}$.
\end{lemma} 

%It follows from Remark~\ref{rem:WhenRM-PCIsIPC} that RM-PC protocols belong to the class of PC protocols. Therefore, Corollary~\ref{lem:RM-PC_NS} is a direct consequence of \cite[Theorem 5.6.2]{Sandholm2010Population-Game}, which states that every PC protocol is Nash stationary.

\subsubsection{Conditions for $\delta$-passivity: Main Theorem and Analysis}
We now proceed to determine conditions for which a RM-PC protocol is $\delta$-passive. Inspired by the Lyapunov and storage functions introduced respectively in~\cite{Hofbauer2007Evolution-in-ga} and~\cite{Fox2013Population-Game}, we choose the $\delta$-storage function we proceed to describe. Given a population $r\in\{1,\dots,\rho\}$ with a protocol $\mathcal{T}^r$ of the RM-PC class, we set our $\delta$-storage function to be $(\mathcal{S}^{\text{\tiny{RM-PC}}})^r:\mathbb{X}^r\times\mathbb{R}^{n^r}\to\mathbb{R}_{\geq 0}$ specified below:
\begin{equation}
\label{eq:SRM-PC}
(\mathcal{S}^{\text{\tiny RM-PC}})^r(x^r,p^r) := \sum_{i=1}^{n^r} \frac{1}{\bar{\tau}^r}\lambda^r_i x^r_i \left(  \sum_{k=1}^{n^r} \psi^r_k(p^r_k - p^r_i) \right)
\end{equation}
where for all $k,i\in \{1,\dots,n^r\}$,  $\psi^r_k:\mathbb{R}^{n^r}\to \mathbb{R}$ is defined as
\begin{equation*}
\psi^r_k(p^r_k - p^r_i) := \int_0^{p^r_k-p^r_i} \phi^r_k(s)ds
\end{equation*}
Denoting $ \sum_{k=1}^{n^r} \psi^r_k(p^r_k - p^r_i)$ by $\gamma^r_i(p^r)$ we can write $(\mathcal{S}^{\text{\tiny{RM-PC}}})^r$ in a more compact form as
\begin{align*}
(\mathcal{S}^{\text{\tiny{RM-PC}}})^r(x^r,p^r) = \sum_{i=1}^{n^r} \frac{1}{\bar{\tau}^r}\lambda^r_i x^r_i \gamma^r_i(p^r)
\end{align*}

The following is the main result of this section.
\vspace{.1 in}
\begin{theorem}
\label{thm:RM-PC_diss} Given $r$ in $\{1,\ldots,\rho\}$, consider that the $r$-th population follows an RM-PC protocol specified by a given $\phi^r$ and a worst-case revision rate ratio~$\lambda_R^r$~(see~(\ref{eq:WorstRatioDefinition})).  The RM-PC protocol for population $r$ is $\delta$-passive if {\bf (i)}~$n^r=2$ \underline{or} {\bf (ii)}~$n^r \geq 3$ and the following inequality holds:
\begin{equation}
\label{cond:RM-PC}
\lambda_R^r < \bar{\lambda}_{\phi^r}(n^r)
\end{equation}
where $\bar{ \lambda}_{\phi^r}$ is determined from $\phi^r$ as follows:
\begin{subequations}
\label{ineq:RM-PC}
\begin{equation}\label{ineq:RM-PC-a}
\bar{ \lambda}_{\phi^r}(n^r):=\min_{1 \leq k\leq n^r} \inf_{p^r\in\mathbb{R}^{n^r}} \left\{\frac{\gamma^r_{k}(p^r)\sum_{i=1}^{n^r}\phi^r_i(p^r_i-p^r_k)}{\sum_{i=1}^{n^r}\phi^r_i(p^r_i-p^r_k)\gamma^r_i(p^r)}\right\}
\end{equation} Although (to avoid cumbersome notation) we do not explicitly indicate in~(\ref{ineq:RM-PC-a}), the infimum  is computed subject to the following constraint on $p^r$:
\begin{equation}
\sum_{i=1}^{n^r}\phi^r_i(p^r_i-p^r_k)\gamma^r_i(p^r) \neq 0
\label{eq:ConstraintForLambdaRMPC}
\end{equation} 
\end{subequations}
\end{theorem}
\vspace{.1in}

In Appendix~\ref{pf:RM-PC_diss} we will prove Theorem~\ref{thm:RM-PC_diss} by showing that $(\mathcal{S}^{\text{\tiny{RM-PC}}})^r$ satisfies~(\ref{eq:PWiseDPass}).

\begin{Remark}{\bf (When to compute~(\ref{ineq:RM-PC}))} According to Theorem~\ref{thm:RM-PC_diss}, an RM-PC protocol is always $\delta$-passive for a population with two strategies, irrespective of the revision rates. Hence, only when $n^r \geq 3$ will one need to compute~(\ref{ineq:RM-PC}) to test whether~(\ref{cond:RM-PC}) holds.
\end{Remark}

Below, we will state a proposition (proved in Appendix~\ref{pf:RM-PCBound>1}) that introduces a simple lower bound for $\bar{ \lambda}_{\phi^r}(n^r)$ that is valid for RM-PC protocols satisfying the following assumption for population $r$.
\begin{Assumption}
\label{assump:prop1}
There is a non-decreasing function ${\bar{\phi}^r:\mathbb{R} \rightarrow \mathbb{R}_{\geq 0}}$ such that the following holds:
    \begin{equation}
        \label{eq:cond1prop1}
        \phi^r_i(\tilde{p}) = \bar{\phi}^r(\tilde{p}), \quad \tilde{p} \in \mathbb{R}, \ i \in \{1,\ldots,n^r\} 
    \end{equation}
\end{Assumption}
\vspace{.1in}
\begin{proposition}
\label{prop:RM-PCBound>1}
Consider that a population $r$ in $\{1,\dots,\rho\}$ (with $n^r\geq 3$) follows an RM-PC protocol. If the protocol satisfies Assumption~\ref{assump:prop1} then the following holds for $n^r \geq 3$:
\begin{equation}
\label{eq:LambdaRMSmithLowerBound}
\bar{\lambda}_{\phi^r} (n^r) \geq \frac{n^r-1}{n^r-2}
\end{equation}

%\begin{equation}
%\label{eq:LambdaRMSmithLowerBound}
%\bar{\lambda}_{\phi^r} (n^r)=\inf_{ p_1\geq \cdots \geq p_{n^r} } \frac{\gamma^r_{n^r}(p^r)\sum_{i=1}^{n^r}\phi^r_i(p^r_i-p^r_{n^r})}{\sum_{i=1}^{n^r}\phi^r_i(p^r_i-p^r_{n^r})\gamma^r_i(p^r)} \geq \frac{n^r-1}{n^r-2}
%\end{equation} where the infimum is computed subject to~(\ref{eq:ConstraintForLambdaRMPC}).
\end{proposition}
\vspace{.1 in}

The proposition's proof given in Appendix~\ref{pf:RM-PCBound>1} introduces an alternative way to compute $\bar{\lambda}_{\phi^r} (n^r)$ for the case in which Assumption~\ref{assump:prop1} holds~(see~(\ref{exp:RM-EPT_diss_equiv_simp})). We make use of (\ref{exp:RM-EPT_diss_equiv_simp}) to simplify our computation of $\bar{\lambda}_{\phi^r}$ for the RM-Smith protocol in \S\ref{subsec:Numerical}.

\vspace{.1in}
\begin{Remark} We can conclude from~(\ref{eq:LambdaRMSmithLowerBound}) that, for the protocols satisfying the conditions of Proposition~\ref{prop:RM-PCBound>1}, $\bar{\lambda}_{\phi^r} (n^r)$ is strictly greater than $1$, which, according to Theorem~\ref{thm:RM-PC_diss}, affords some $\delta$-passivity robustness with respect to $\lambda_R^r$ regardless of the number of strategies. This fact is in contrast to previous results establishing $\delta$-passivity only for protocols in which $\lambda_R^r$ was exactly $1$ (see~Remark~\ref{rem:WhenRM-PCIsIPC}).
\end{Remark}

The following counterexample illustrates why we need Assumption~\ref{assump:prop1} in Proposition~\ref{prop:RM-PCBound>1}.

\begin{Counterexample}
Consider that $n^r=3$ and population $r$ adopts an RM-PC protocol specified by $\phi^r_1(\cdot) = [\cdot]_+^2$, $\phi^r_2(\cdot) = \phi^r_3(\cdot) = [\cdot]_+$. This protocol violates Assumption~\ref{assump:prop1} and, as we proceed to show, it will infringe~(\ref{eq:LambdaRMSmithLowerBound}) with~$\bar{\lambda}_{\phi^r}(3) = 1$. To do so, consider the following inequality that we obtain by using $p^r_1 = 0$, $p^r_2 = -\epsilon$, $p^r_3 = -\epsilon+\epsilon^{7/4}$, with $\epsilon>0$, when computing the infimum in (\ref{ineq:RM-PC-a}):
\begin{equation*}
\bar{\lambda}_{\phi^r}(3) \leq \lim_{\epsilon \rightarrow 0^+} \frac{(2\epsilon^3+3\epsilon^{7/2})(\epsilon^2+\epsilon^{7/4})}{2(\epsilon-\epsilon^{7/4})^3\epsilon^{7/4}} = 1
\end{equation*}
\end{Counterexample}
\vspace{.1 in}

\subsection{Numerical Evaluation of $\bar{\lambda}_{\phi^r}$ for RM-Smith}
\label{subsec:Numerical}

\begin{figure}[h!]
    \centering
    \begin{tikzpicture}[scale=0.8]
        \begin{axis}[
            ymode=log,
            log basis y=10,
            ytick=\empty,
            %log ticks with fixed point,
            % for log axes, x filter operates on LOGS.
            % and log(x * 1000) = log(x) + log(1000):
            %y filter/.code=\pgfmathparse{#1 + 6.90775527898214},
            width=0.5\textwidth,
            height=0.3\textwidth,
            ylabel style={font=\color{white!15!black}},
            ylabel={Logarithmic scale},
            xlabel style={font=\color{white!15!black}},
            xlabel={Number of strategies ($n^r$)}]
            \pgfplotsset{compat=1.10}
            \pgfplotstableread{
            3	0.3010299957
            4	0.1760912591
            5	0.1249387366
            6	0.09691001301
            7	0.07918124605
            8	0.0669467896
            9	0.05799194698
            10	0.05115252245
            }\datatable
            % \pgfplotstableread{results.dat}\datatable
            \pgfplotstablegetrowsof{\datatable}
            \addplot+[mark=*, only marks, forget plot] table {\datatable};
            \addplot+[only marks] table  {\datatable}
            % pos = (index-1)/(N-1)  (index starting from 1)
            node[pos=0, pin={[pin distance = 1 mm]85:2}]{}
            node[pos=1/7, pin={[pin distance = 1 mm]90:1.5}]{}
            node[pos=2/7, pin={[pin distance = 1 mm]90:1.33}]{}
            node[pos=3/7, pin={[pin distance = 1 mm]90:1.25}]{}
            node[pos=4/7, pin={[pin distance = 1 mm]90:1.2}]{}
            node[pos=5/7, pin={[pin distance = 1 mm]90:1.16}]{}
            node[pos=6/7, pin={[pin distance = 1 mm]90:1.14}]{}
            node[pos=1, pin={[pin distance = 1 mm]90:1.12}]{}
            ;
            %\addlegendentry{\footnotesize $(n-1)/(n-2)$}
            \pgfplotstableread{
            3	0.9751347469
            4	0.7704734633
            5	0.6675681373
            6	0.602059993
            7	0.5553833492
            8	0.5198160661
            9	0.4914733949
            10	0.4681524677
            }\datatable
            % \pgfplotstableread{results.dat}\datatable
            \pgfplotstablegetrowsof{\datatable}
            \addplot+[mark=*, only marks, forget plot] table {\datatable};
            \addplot+[only marks] table  {\datatable}
            % pos = (index-1)/(N-1)  (index starting from 1)
            node[pos=0, pin={[pin distance = 1 mm]-88:9.44}]{}
            node[pos=1/7, pin={[pin distance = 1 mm]-90:5.89}]{}
            node[pos=2/7, pin={[pin distance = 1 mm]-90:4.65}]{}
            node[pos=3/7, pin={[pin distance = 1 mm]-90:4}]{}
            node[pos=4/7, pin={[pin distance = 1 mm]-90:3.59}]{}
            node[pos=5/7, pin={[pin distance = 1 mm]-90:3.3}]{}
            node[pos=6/7, pin={[pin distance = 1 mm]-90:3.1}]{}
            node[pos=1, pin={[pin distance = 1 mm]-90:2.93}]{}
            ;
            %\addlegendentry{\footnotesize $\bar{\phi}_{[\cdot]_+}$}
        \end{axis};
        \node[anchor=south,black!50!red] at (6.3,3) {\large $\bar{\lambda}_{\text{\tiny RM-Smith}} (n^r)$};
        \node[anchor=south,black!50!blue] at (1.3,0.3) {\Large $\frac{n^r-1}{n^r-2}$};
    \end{tikzpicture}
    \caption{Comparing $\bar{\lambda}_{\text{\tiny RM-Smith}}$ with the lower-bound in~(\ref{eq:LambdaRMSmithLowerBound}).}
    \label{fig:SmithBounds}
\end{figure}
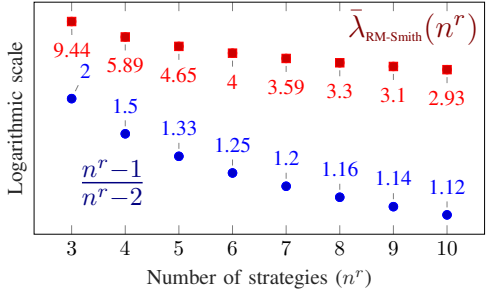

We start by denoting $\bar{\lambda}_{\phi^r}$ for the RM-Smith protocol as $\bar{\lambda}_{\text{\tiny RM-Smith}}$, which we determine by computing~(\ref{ineq:RM-PC-a}) numerically \cite{semih_kara_2021_5068102}. In Fig.~\ref{fig:SmithBounds}, we plot $\bar{\lambda}_{\text{\tiny RM-Smith}}$ and the lower bound in (\ref{eq:LambdaRMSmithLowerBound}) for ${1 \leq n^r \leq 10}$. Notice that since the RM-Smith protocol satisfies Assumption~\ref{assump:prop1}, the lower bound in (\ref{eq:LambdaRMSmithLowerBound}) holds for $\bar{\lambda}_{\text{\tiny RM-Smith}}$, for any $n^r \geq 3$. 
%From here we observe that $\tfrac{n^r-1}{n^r-2}$ is indeed a lower-bound to $\bar{\lambda}^r_{\text{\tiny RM-Smith}} (n^r)$.

The plots in Fig.~\ref{fig:SmithBounds} illustrate that the lower-bound in~(\ref{eq:LambdaRMSmithLowerBound}) may be conservative \textendash~a feature of it being valid for a large subclass of RM-PC protocols. Notably, from the values of $\bar{\lambda}_{\text{\tiny RM-Smith}}$ plotted in Fig.~\ref{fig:SmithBounds} we observe that the RM-Smith protocol satisfies (\ref{cond:RM-PC}) even if the revision rates of the $r$-th population vary by a multiplicative factor \underline{exceeding $9$} when $n^r=3$. For $n^r=10$, the revision rates of the $r$-th population is allowed to vary by a multiplicative factor of \underline{nearly $3$}.

\section{Establishing GAS Of The Equilibria}
\label{subsec:RM-EPTPCGAS}

We proceed to use Lemma~\ref{lem:RM-PC_NS} and Theorem~\ref{thm:RM-PC_diss} in conjunction with~\cite{Arcak2020Dissipativity-T} to draw conclusions about the equilibrium stability of the mean closed loop.

\subsection{GAS For Memoryless $\mathcal{F}$}
Given a payoff described by a game $\mathcal{F}$ (memoryless), the following theorem establishes conditions for GAS of $\mathbb{NE}(\mathcal{F})$ under the mean closed loop (\ref{eq:closedloopmodelF}) formed by $\mathcal{F}$ and an EDM whose populations follow RM-PC protocols. We state the theorem without proof because it follows directly from \cite[Corollary 1]{Arcak2020Dissipativity-T} in conjunction with Lemma~\ref{lem:RM-PC_NS} and Theorem~\ref{thm:RM-PC_diss}.

\vspace{.1 in}
\begin{theorem}
\label{cor:RM-EPTGAS}
Consider that a game $\mathcal{F}$ is given and that each population follows an RM-PC protocol. If the protocols satisfy the conditions of Theorem~\ref{thm:RM-PC_diss} and the game is weighted-contractive  (see Assumption~\ref{ass:WeightedContractiveF}), then $\mathbb{NE}(\mathcal{F})$ is a GAS equilibria set of (\ref{eq:closedloopmodelF}).
\end{theorem}
\vspace{.1 in}

Notice that Theorem~\ref{cor:RM-EPTGAS} generalizes~\cite[Theorem~7.1]{Hofbauer2009Stable-games-an} in two ways. In comparison to the latter, which presumes that the game is contractive and the revision rates are identical within each population, the former allows for weighted-contractive games and it contends with unequal revision rates so long as they satisfy the conditions of the theorem. The stability theorems in~\cite{Arcak2020Dissipativity-T} allow for weighted-contractive games but the article lacks the results needed to consider the case in which the revision rates within each population are different.

\subsection{GAS For PDM}
The following theorem is the counterpart of Theorem~\ref{cor:RM-EPTGAS} for the case in which the payoff mechanism is a PDM. We state it without proof as the theorem follows directly from~\cite[Theorem~2]{Arcak2020Dissipativity-T}, with $\Pi$ selected as in~\cite[(8)]{Arcak2020Dissipativity-T}, in conjunction with Lemma~\ref{lem:RM-PC_NS} and Theorem~\ref{thm:RM-PC_diss}.

\vspace{.1 in}
\begin{theorem}
\label{cor:RM-EPTGASPDM}
Consider that a PDM is given and that each population follows an RM-PC protocol. If the protocols satisfy the conditions of Theorem~\ref{thm:RM-PC_diss} and the PDM satisfies Assumption~\ref{ass:PDMDissipative}, then the equilibria set of~(\ref{eq:closedloopmodelPDM}) is GAS. In addition, $\mathbb{NE}(\mathcal{F}_{\mathcal{G},\mathcal{H}})$ are the $x$ components of the equilibria.
\end{theorem}
\vspace{.1 in}

The theorem above exemplifies how our results can extend the applicability of~\cite[Theorem~2]{Arcak2020Dissipativity-T} to the case in which each population follows an RM-PC protocol.

\subsection{Generalizing Theorems~\ref{cor:RM-EPTGAS}
and~\ref{cor:RM-EPTGASPDM}} 
\label{subsec:GeneralizationThms2nad3}
It is useful for exploring possible generalizations of Theorems~\ref{cor:RM-EPTGAS} and~\ref{cor:RM-EPTGASPDM} to observe that they remain valid for any protocol satisfying~(\ref{eq:PWiseDPass}). For instance, we could have stated Theorems~\ref{cor:RM-EPTGAS} and~\ref{cor:RM-EPTGASPDM} more generally by requiring that each population follows either an RM-PC protocol satisfying the conditions of Theorem~\ref{thm:RM-PC_diss} or a so-called {\it excess payoff target}~(EPT) protocol~\cite{Sandholm2005Excess-payoff-d} whose $\delta$-passivity is stated in \cite[Theorem~4.4]{Fox2013Population-Game}, and discussed more generally in~\cite[\S VI.B]{Park2019From-Population} and references therein. It should be noted that, in such theorems, the EPT protocol would not be rate-modified (hindering its applicability in our context of strategy-dependent revision rates), which justifies our decision to not commit space to proving it rigorously here. (See future directions in~\S \ref{sec:conclusions}.)

\section{Numerical Examples}

As industrial-grade data-driven processing centers and vehicle to vehicle networks are becoming more prevalent, life cycles of DRAMs used in these applications emerge as important benchmarks. To provide examples of how our results can come into play, we look into the  HPG and its smoothed version, introduced respectively in Example \ref{ex:Game} and Appendix~\ref{app:ExampleSmoothedExample}, in the context of the DRAM market.

\subsection{A DRAM Market HPG} 
\label{subsec:has_vs_price_in_DRAM}

We proceed by introducing an HPG in the context of the DRAM market. There are two populations, each representing a class of systems in which DRAMs are commonly used. Namely, classes~1 and~2 are respectively industrial and automotive systems. We assume that there are 3 manufacturers producing DRAMs with failure rates in these utilization classes given by $\lambda_1^1 = 5$, $\lambda_1^2=10$, $\lambda_2^1 = 4$, $\lambda_2^2=9$ and $\lambda_3^1 = 3$, $\lambda_3^2=5$, where $\lambda_i^j$ is the failure rate of DRAMs produced by manufacturer $i$ when utilized in class $j$. Moreover, we assume that the replacement costs for industrial and automotive DRAMs are $\beta^1=2$ and $\beta^2=1$, respectively.

We assume that the component price from manufacturer $i\in\{1,2,3\}$, which is the $\mathcal{C}_i$ in \eqref{eq:examplepayoff}, is determined as the sum of a fixed production cost, $\mathcal{C}_{0i}$, and a term reflecting the pull-back inflation, $\mathcal{C}_{pi}$. In order to reflect the pull-back inflation on the cost we will use a quadratic term given by $\mathcal{C}_{pi}(\mathcal{D}_i(x)) = (\mathcal{D}_i(x))^2 = (\alpha^1x^1_i+\alpha^2x^2_i)^2$, where $\alpha^r$ is in proportion to the share of class-$r$ in the DRAM market. Finally, we set $\alpha^1=1$ and $\alpha^2=2$, and the fixed DRAM production costs to be $\mathcal{C}_{01}=1$, $\mathcal{C}_{02}=1.2$ and $\mathcal{C}_{03}=1.5$, which completes the construction of $\mathcal{F}$, as in~(\ref{eq:examplepayoff}), for our DRAM market HPG.

\noindent {\bf Note:} We would like to clarify that the functions and parameters selected in this section are for illustration purposes, and they are not estimated from data.

\subsection{Dynamics Under the RM-Smith Protocol} 
\label{subsec:has_vs_price_dyn}

Now we describe how Theorem~\ref{cor:RM-EPTGAS} can be utilized. Consider the mean closed loop~(\ref{eq:closedloopmodelF}) with $\mathcal{F}$ constructed in~\S\ref{subsec:has_vs_price_in_DRAM} and the RM-Smith EDM~(\ref{eq:SmithEDMDef}). Assume that initially the buyers are distributed on the manufacturers according to $x^1(0) = x^2(0) = (2/3,1/6,1/6)$.

Since the failure rates satisfy the condition of Theorem~\ref{thm:RM-PC_diss} and $\mathcal{F}$ satisfies Assumption~\ref{ass:WeightedContractiveF}, we can invoke Theorem~\ref{cor:RM-EPTGAS} to conclude that $x$ converges to $\mathbb{NE}(\mathcal{F})$, which in this case is the singleton $(x^1)^* = (0,1,0)$, $(x^2)^* = (0,0,1)$ \cite{semih_kara_2021_5068102}. For this example, the trajectory and the time domain plot of $x$ are portrayed respectively in Fig.~\ref{fig:DRAM_Smith} and Fig.~\ref{fig:DRAM_Smith_time}. 
%From the definition of the Nash equilibria of a population game, presented in \S\ref{subsec:NE-Intro}, we can verify that the limit point of the mean social state depicted in Fig~\ref{fig:DRAM_Smith}, given by $(x^1)^* = (0,1,0)$, $(x^2)^* = (0,0,1)$, is indeed a Nash equilibrium of the HPG for the DRAM market with the selected parameters.

% We note that a characterization of Nash equilibria of a population game $\mathcal{F}$ is $x\in\mathbb{NE}(\mathcal{F})$ if and only if for all $r\in\{1,\dots,\rho\}$, $x^r_i > 0$ implies $i\in\arg\max_{j\in\{1,\dots,n^r\}}\mathcal{F}^r_j(x)$. From this characterization we can easily check that the limit point of the mean social state depicted in Fig~\ref{fig:DRAM_Smith}, given by $(x^1)^* = (0,1,0)$, $(x^2)^* = (0,0,1)$, is indeed a Nash equilibrium of the hassle vs. price game for the DRAM market with the introduced parameters.

\tikzset{decorated arrows/.style={
		postaction={
			decorate,
			decoration={
				markings,
				mark = between positions 0.1 and 1 step 9mm with {\arrow[black]{stealth};}
			}
		}
	},
}
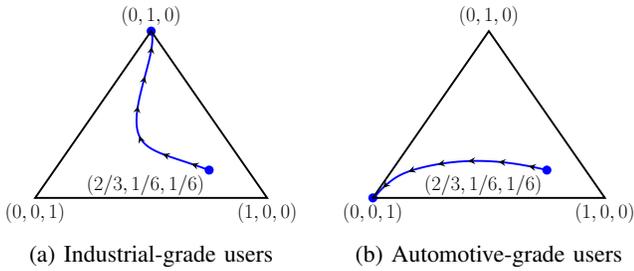
\begin{figure}
    \begin{subfigure}{0.49\columnwidth}
    	\centering
		\begin{tikzpicture}[scale=0.45]
    		\tikzstyle{every node}=[font=\LARGE]
    		\begin{axis}[
        		axis y line=center,
        		axis x line=middle, 
        		axis on top=true,
        		xmin=0,
        		xmax=1,
        		ymin=0,
        		ymax=1,
        		clip=false,
        		axis line style={draw=none},
        		tick style={draw=none},
        		xticklabels={,,},
        		yticklabels={,,},
        		ultra thick
        		] 
        		\addplot[ultra thick, blue, decorated arrows] table [col sep=comma, mark=none] {data1_Smith.txt};
        		\draw [blue, fill=blue] (0.75,0.14434) circle (3pt);
        		\draw [blue, fill=blue] (0.5,0.86603) circle (3pt);
        		\draw (0.75,0.14434) node[anchor=north east]{$(2/3,1/6,1/6)$};
        		\draw (0,0) node[anchor=north]{$(0,0,1)$}
        		-- (1,0) node[anchor=north]{$(1,0,0)$}
        		-- (1/2,1.7320508076/2) node[anchor=south]{$(0,1,0)$}
        		-- cycle;
    		\end{axis}
		\end{tikzpicture}
		\caption{Industrial-grade users}
    \end{subfigure}
    \hfil
	\begin{subfigure}{0.49\columnwidth}
    	\centering
    	\begin{tikzpicture}[scale=0.45]
        	\tikzstyle{every node}=[font=\LARGE]
        	\begin{axis}[
            	axis y line=center,
            	axis x line=middle, 
            	axis on top=true,
            	xmin=0,
            	xmax=1,
            	ymin=0,
            	ymax=1,
            	clip=false,
            	axis line style={draw=none},
            	tick style={draw=none},
            	xticklabels={,,},
            	yticklabels={,,},
            	ultra thick
            	] 
            	\addplot[ultra thick, blue, decorated arrows] table [col sep=comma, mark=none] {data2_Smith.txt};
            	\draw [blue, fill=blue] (0.75,0.14434) circle (3pt);
            	\draw [blue, fill=blue] (0,0) circle (3pt);
            	\draw (0.75,0.14434) node[anchor=north east]{$(2/3,1/6,1/6)$};
            	\draw (0,0) node[anchor=north]{$(0,0,1)$}
            	-- (1,0) node[anchor=north]{$(1,0,0)$}
            	-- (1/2,1.7320508076/2) node[anchor=south]{$(0,1,0)$}
            	-- cycle;
        	\end{axis}
    	\end{tikzpicture}
    	\caption{Automotive-grade users}
    \end{subfigure}
	\caption{Trajectory of distribution of DRAM buyers on manufacturers under the HPG and RM-Smith protocol.}
	\label{fig:DRAM_Smith}
\end{figure}

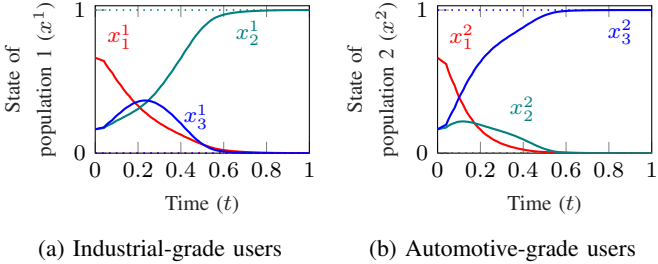
\begin{figure}
\hspace{-0.2 in}
    \begin{tabular}[b]{cc}
    	\multicolumn{2}{c}{
    	\begin{subfigure}[b]{0.5\columnwidth}
			\centering
			\begin{tikzpicture}[scale=1]
			    \tikzstyle{every node}=[font=\footnotesize]
			    \begin{axis}[
    				width=\textwidth,
    				height=0.8\textwidth,
    				xmin=0,
    				xmax=1,
    				ymin=0,
    				ymax=1.03,
    				ylabel style={font=\color{white!15!black}, align=center},
    				ylabel={\footnotesize State of\\\footnotesize population 1 ($x^1$)},
    				xlabel style={font=\color{white!15!black}},
    				xlabel={\footnotesize Time ($t$)},
    				ytick={0,1}]
    				\addplot[thick, red] table [col sep=comma, mark=none] {x11.txt} node[above right,pos=0.01] { $x_1^1$};
    				\addplot[thick, teal] table [col sep=comma, mark=none] {x12.txt} node[below,pos=0.8] { $x_2^1$};
    				\addplot[thick, blue] table [col sep=comma, mark=none] {x13.txt} node[right,pos=0.4] { $x_3^1$};
    				\draw [semithick, red, dotted] (0,0) -- (1,0);
    				\draw [semithick, teal, dotted] (0,1) -- (1,1);
    				\draw [semithick, blue, dotted] (0,0) -- (1,0);
			    \end{axis}
    	    \end{tikzpicture}
    	    \caption{Industrial-grade users}
    	    \label{fig:x1}
    	\end{subfigure}
    	\begin{subfigure}[b]{0.5\columnwidth}
    		\centering
    		\begin{tikzpicture}[scale=1]
    		    \tikzstyle{every node}=[font=\footnotesize]
    			\begin{axis}[
    				width=\textwidth,
    				height=0.8\textwidth,
    				xmin=0,
    				xmax=1,
    				ymin=0,
    				ymax=1.03,
    				ylabel style={font=\color{white!15!black}, align=center},
    				ylabel={\footnotesize State of\\\footnotesize population 2 ($x^2$)},
    				xlabel style={font=\color{white!15!black}},
    				xlabel={\footnotesize Time ($t$)},
    				ytick={0,1}]
    				\addplot[thick, red] table [col sep=comma, mark=none] {x21.txt} node[above right,pos=0.01] { $x_1^2$};
    				\addplot[thick, teal] table [col sep=comma, mark=none] {x22.txt} node[above right,pos=0.3] { $x_2^2$};
    				\addplot[thick, blue] table [col sep=comma, mark=none] {x23.txt} node[below,pos=0.9] { $x_3^2$};
    				\draw [semithick, red, dotted] (0,0) -- (1,0);
    				\draw [semithick, teal, dotted] (0,0) -- (1,0);
    				\draw [semithick, blue, dotted] (0,1) -- (1,1);
    			\end{axis}
    		\end{tikzpicture}
    		\caption{Automotive-grade users}
    		\label{fig:x2}
    	\end{subfigure}}
    \end{tabular}
    \caption{Time domain plots of the distribution of DRAM buyers on manufacturers under the HPG and RM-Smith protocol.}
    \label{fig:DRAM_Smith_time}
\end{figure}

\subsection{Smoothed HPG for the DRAM Market and Dynamics Under the RM-Smith Protocol} 

% In this section we illustrate an application of our results on mean closed loops with PDMs. Consider a mean closed loop formed by a PDM given by the smoothed hassle vs. price game introduced in Appendix~\ref{app:ExampleSmoothedExample} and an EDM induced by the RM-Smith protocol. In the smoothed hassle vs. price game, we set the time constant $a$ to be $25$ and take all other parameters to be as given in~\S\ref{subsec:has_vs_price_in_DRAM}. Moreover we assume the initial state of the PDM to be $q(0)=(0,0,0)$ and initially all buyers to be distributed on the manufacturers according to the initial distribution given in~\S\ref{subsec:has_vs_price_dyn}, that is $x^1(0)=x^2(0)=(2/3,1/6,1/6)$.

We also carried out an analysis that is analogous to that in \S\ref{subsec:has_vs_price_dyn}, but for the mean closed loop~(\ref{eq:closedloopmodelPDM}) with the RM-Smith EDM~(\ref{eq:SmithEDMDef}) and the smoothed HPG PDM specified in Appendix~\ref{app:ExampleSmoothedExample}. We selected $a=5$ in \eqref{eq:sm_hassle_vs_price} and we kept all the other parameters unchanged from \S\ref{subsec:has_vs_price_dyn}.

\begin{figure}
\hspace{-0.2 in}
    \begin{tabular}[b]{cc}
    	\multicolumn{2}{c}{
    	\begin{subfigure}[b]{\columnwidth}
    		\centering
    		\begin{tikzpicture}[scale=1]
    			\tikzstyle{every node}=[font=\footnotesize]
    			\begin{axis}[
    				width=\textwidth,
    				height=0.4\textwidth,
    				xmin=0,
    				xmax=1,
    				ymin=0,
    				ymax=5.9,
    				ylabel style={font=\color{white!15!black}},
    				ylabel={\footnotesize PDM state ($q$)},
    				xlabel style={font=\color{white!15!black}},
    				xlabel={\footnotesize Time ($t$)},
    				ytick={1,2.1999,5.5},
    				yticklabels={1,$2.1\overline{9}$,5.5}]
    				\addplot[thick, red] table [col sep=comma, mark=none] {v_q1.txt} node[above,pos=0.9] {$q_1$};
    				\addplot[thick, blue] table [col sep=comma, mark=none] {v_q2.txt} node[above,pos=0.9] {$q_2$};
    				\addplot[thick, teal] table [col sep=comma, mark=none] {v_q3.txt} node[below,pos=0.9] {$q_3$};
    				\draw [semithick, red, dotted] (0,1) -- (1,1);
    				\draw [semithick, blue, dotted] (0,2.1999) -- (1,2.1999);
    				\draw [semithick, teal, dotted] (0,5.5) -- (1,5.5);
    			\end{axis}
    		\end{tikzpicture}
    		\caption{State of the smoothed HPG}
    		\label{fig:PDM_q}
    	\end{subfigure}}\\\noindent
    	\begin{subfigure}[b]{0.5\columnwidth}
    		\centering
    		\begin{tikzpicture}[scale=1]
    			\tikzstyle{every node}=[font=\footnotesize]
    			\begin{axis}[
    				width=\textwidth,
    				height=0.8\textwidth,
    				xmin=0,
    				xmax=1,
    				ymin=0,
    				ymax=1.03,
    				ylabel style={font=\color{white!15!black}, align=center},
    				ylabel={\footnotesize State of\\\footnotesize population 1 ($x^1$)},
    				xlabel style={font=\color{white!15!black}},
    				xlabel={\footnotesize Time ($t$)},
    				ytick={0,1}]
    				\addplot[thick, red] table [col sep=comma, mark=none] {v_x11.txt} node[above, pos=0.1] {$x_1^1$};
    				\addplot[thick, teal] table [col sep=comma, mark=none] {v_x12.txt} node[below,pos=0.8] {$x_2^1$};
    				\addplot[thick, blue] table [col sep=comma, mark=none] {v_x13.txt} node[right,pos=0.5] {$x_3^1$};
    				\draw [semithick, red, dotted] (0,0) -- (1,0);
    				\draw [semithick, teal, dotted] (0,1) -- (1,1);
    				\draw [semithick, blue, dotted] (0,0) -- (1,0);
    			\end{axis}
    		\end{tikzpicture}
    		\caption{Industrial-grade users}
    		\label{fig:PDM_x1}
    	\end{subfigure} &
    	\begin{subfigure}[b]{0.5\columnwidth}
            \centering
    		\begin{tikzpicture}[scale=1]
    			\tikzstyle{every node}=[font=\footnotesize]
    			\begin{axis}[
    				width=\textwidth,
    				height=0.8\textwidth,
    				xmin=0,
    				xmax=1,
    				ymin=0,
    				ymax=1.03,
    				ylabel style={font=\color{white!15!black}, align=center},
    				ylabel={\footnotesize State of\\\footnotesize population 2 ($x^2$)},
    				xlabel style={font=\color{white!15!black}},
    				xlabel={\footnotesize Time ($t$)},
    				ytick={0,1}]
    				\addplot[thick, red] table [col sep=comma, mark=none] {v_x21.txt} node[above, pos=0.05] {$x_1^2$};
    				\addplot[thick, teal] table [col sep=comma, mark=none] {v_x22.txt} node[above right,pos=0.35] {$x_2^2$};
    				\addplot[thick, blue] table [col sep=comma, mark=none] {v_x23.txt} node[below,pos=0.9] {$x_3^2$};
    				\draw [semithick, red, dotted] (0,0) -- (1,0);
    				\draw [semithick, teal, dotted] (0,0) -- (1,0);
    				\draw [semithick, blue, dotted] (0,1) -- (1,1);
    			\end{axis}
    		\end{tikzpicture}
    		\caption{Automotive-grade users}
    		\label{fig:PDM_x2}
    	\end{subfigure}
    \end{tabular}
    \caption{Time domain plots of the PDM's state and distribution of DRAM buyers on manufacturers under the smoothed HPG and RM-Smith protocol.}
    \label{fig:PDM}
\end{figure}
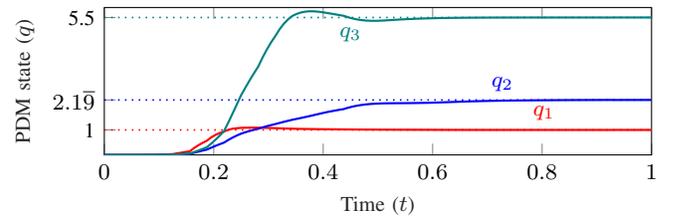
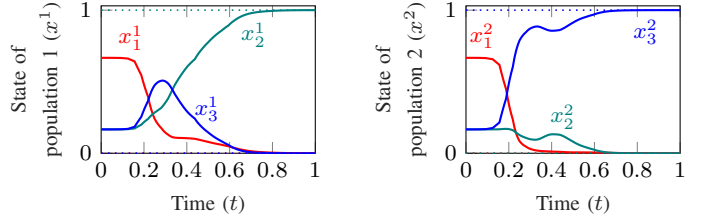

Since the failure rates satisfy the condition of Theorem~\ref{thm:RM-PC_diss}, we can invoke Remark~\ref{rem:ThePDMExampleWorks} to conclude from Theorem~\ref{cor:RM-EPTGASPDM} and Remark~\ref{rem:FisFGH}~(Appendix~\ref{app:ExampleSmoothedExample}) that, like in \S\ref{subsec:has_vs_price_dyn}, $x$ will converge to $\mathbb{NE}(\mathcal{F})$. The time evolution of the PDM's state $q$ and the social state $x$ are plotted in Fig.~\ref{fig:PDM}, indicating that $x^1$ and $x^2$ indeed converge respectively to $(0,1,0)$ and $(0,0,1)$.

% Moreover, $x$ components of the elements of this equilibria set is equal to $\mathbb{NE}(\mathcal{F_{G,H}})$, where plugging $\dot{q}(t)=0$ to (\ref{eq:sm_hassle_vs_price}) reveals that for all $r\in\{1,\dots,\rho\}$ and $i\in\{1,\dots,n^r\}$
% \begin{align*}
% (\mathcal{F_{G,H}})_i^r(x) = -\beta^r_i\lambda_i^r - \mathcal{C}_i(\mathcal{D}_i(x))
% \end{align*}
% that is, the hassle vs. price game is the stationary game of its smoothed modification. As a result, $x$ components of the elements of the equilibria set for the mean closed loop coincides with $\mathbb{NE}(\mathcal{F})$, where $\mathcal{F}$ here represents the hassle vs. price game with the parameters given in~\S\ref{subsec:has_vs_price_in_DRAM}.

% For this example, time domain plots of the PDM's state and mean social state are portrayed in Fig.~\ref{fig:PDM}. Note from this figure that states of populations $1$ and $2$ converge respectively to $(0,1,0)$ and $(0,0,1)$, which, as discussed in~\S\ref{subsec:has_vs_price_dyn}, belong to the set of Nash equilibria of the hassle vs. price game with parameters given in~\S\ref{subsec:has_vs_price_in_DRAM}. Since the hassle vs. price game and the stationary game of its smoothed modification coincide, we arrive at the conclusion that the limit point of the mean social state depicted in Fig.~\ref{fig:PDM} indeed belongs to $\mathbb{NE}(\mathcal{F_{G,H}})$.

\section{Conclusions and Future Directions}
\label{sec:conclusions}

In this article we were able to generalize the approach in~\cite{Fox2013Population-Game} and~\cite{Arcak2020Dissipativity-T} to a class of pairwise comparison protocols we called RM-PC for which the agents' revision rates may depend on their current strategies. We stated and proved two theorems establishing global asymptotic stability of the equilibria of the mean closed loop for the cases when the payoff mechanism is a memoryless game or a payoff dynamics model (PDM). These results rely on Theorem~\ref{thm:RM-PC_diss} establishing conditions for $\delta$-passivity of the RM-PC protocol. Proposition~\ref{prop:RM-PCBound>1} establishes for an RM-PC protocol sub-class a rather simple (but more conservative) sufficient condition for $\delta$-passivity.

\noindent {\bf Future Direction 1:} Motivated by the discussion in \S\ref{subsec:GeneralizationThms2nad3}, a meaningful next step would be to propose a rate-modified version of the excess payoff target (EPT) protocol whose $\delta$ passivity we would then study by appropriately generalizing the approach \cite{Fox2013Population-Game} and~\cite{Arcak2020Dissipativity-T}.

\noindent {\bf Future Direction 2:} Although Theorem~\ref{thm:RM-PC_diss} guarantees $\delta$-passivity of an RM-PC protocol for any revision rates when there are two strategies (undoubtedly a strong result), if there are three or more strategies it only provides a sufficient condition. Considering that we were unable to construct an example of an RM-PC protocol that is not $\delta$-passive when the condition fails, we believe that it would be important to continue to investigate whether such an example exists or whether the condition could be weakened.

\appendix
\subsection{Smoothed HPG: A PDM Example}
\label{app:ExampleSmoothedExample}

The following is an example of a PDM that can be viewed as a dynamic version of Example~\ref{ex:Game}. Our construction parallels that in~\cite[\S VI.A]{Arcak2020Dissipativity-T}.
\begin{Example}
Given a positive time constant $a$ and parameters as defined in Example~\ref{ex:Game}, the following is the "smoothed\footnote{In~\cite{Fox2013Population-Game}, the authors argue that this type of dynamical modification smooths short-term fluctuations and isolates longer-term trends.}" HPG PDM:
\begin{subequations}
\label{eq:sm_hassle_vs_price}
\begin{align}
a \dot{q}(t) & = -q(t) + \begin{bmatrix} \mathcal{C}_1 \Big ( \mathcal{D}_1 \big( x(t) \big) \Big ) \\ \vdots \\ \mathcal{C}_{\kappa} \Big ( \mathcal{D}_{\kappa} \big( x(t) \big) \Big )
\end{bmatrix}, \ q(0) \in \mathfrak{Q}_0, \ t\geq 0 \\
p^r_i(t) & = -\beta^r \lambda^r_i-  q_i(t), \quad 1 \leq i \leq \kappa, \ 1 \leq r \leq \rho
\end{align}
\end{subequations} Here $\mathfrak{Q} = \mathfrak{Q}_0 := [0,\bar{d}]^\kappa$. We can also specify a set $\mathfrak{P}$ that includes all possible $p$ as follows: $$ \mathfrak{P}:=\left \{  p \in \mathbb{R}^{\kappa\rho} \ \Big |  \ p_i^r = -\beta^r \lambda^r_i-  q_i, \text{ for some $q$ in $\mathfrak{Q}$} \right \}$$
\end{Example}

\vspace{.1 in}
\begin{Remark}\label{rem:FisFGH} It follows immediately from~(\ref{eq:sm_hassle_vs_price}), (\ref{eq:examplepayoff}) and~(\ref{eq:stationarygame}) that, \underline{for the smoothed HPG, $\mathcal{F}$ is identical to $\mathcal{F}_{\mathcal{G},\mathcal{H}}$}.
\end{Remark}

\subsection{Proofs of Theorem \ref{thm:RM-PC_diss} and Proposition \ref{prop:RM-PCBound>1}}
\label{subsec:RM-PCProofs}

Before presenting the proofs of Theorem~\ref{thm:RM-PC_diss} and Proposition~\ref{prop:RM-PCBound>1}, we define a partial order $\succ$ (respectively $\succeq$) on elements of $\mathbb{R}^{n^r}$ as follows. Given any $x,y\in \mathbb{R}^{n^r}$ we write $x\succ y$ (respectively $x \succeq y$) if and only if $x_i > y_i$ (respectively $x_i \geq y_i$) for all $i\in\{1,\dots,n^r\}$. Moreover, given $\lambda^r\in\mathbb{R}^{n^r}$ and $u^r,l^r\in \mathbb{R}$ with $u^r>l^r$, we use a slight abuse of notation and let $u^r\succ \lambda^r \succ l^r$ (respectively $u^r\succeq \lambda^r \succeq l^r$) denote $u^r>\lambda_i^r> l^r$ (respectively $u^r\geq\lambda_i^r\geq l^r$) for all $i\in\{1,\dots,n^r\}$.

\subsubsection{Proof of Theorem \ref{thm:RM-PC_diss}}
\label{pf:RM-PC_diss}

We want to show that the candidate storage function $(\mathcal{S}^{\text{\tiny{RM-PC}}})^r$, given by (\ref{eq:SRM-PC}), satisfies $\delta$-passivity for RM-PC protocols that meet either $n^r=2$ or condition (\ref{cond:RM-PC}). To establish notational convenience, in the rest of the proof we drop the superscript $r$. 

Recall that the component of the EDM corresponding to a population following an RM-PC protocol is given by 
\begin{align*}
\mathcal{V}_i^{\text{\tiny{RM-PC}}} (x,p) &= \sum_{j=1, j \neq i}^{n} \lambda_j \tfrac{1}{\bar{\tau}} \phi_{i} (p_i -p_j) x_j \\ 
&\quad- \sum_{j=1,j \neq i}^{n} \lambda_i \tfrac{1}{\bar{\tau}} \phi_{j} (p_j -p_i) x_i,\quad i\in\{1,\dots,n\}
\end{align*}
and the proposed storage function corresponding to this population is given by
\begin{equation*}
\mathcal{S}^{\text{\tiny RM-PC}}(x,p) = \sum_{i=1}^{n} \frac{1}{\bar{\tau}}\lambda_i x_i \left(  \sum_{k=1}^{n} \psi_k(p_k - p_i) \right)
\end{equation*}
where we note the omission of the $r$ superscript in both expressions.

With our choice of $\delta$-storage function we have
\begin{align*}
&\frac{\partial \mathcal{S}^{\text{\tiny{RM-PC}}}}{\partial x}(x,p)\mathcal{V}^{\text{\tiny{RM-PC}}}(x,p) + \frac{\partial \mathcal{S}^{\text{\tiny{RM-PC}}}}{\partial p}(x,p)u\\
&= \sum_{i=1}^n \frac{1}{\bar{\tau}}\lambda_i \mathcal{V}^{\text{\tiny{RM-PC}}}_i(x,p) \gamma_i(p) +\\
&\quad
\begin{bmatrix}
	u\\
	\mathcal{V}^{\text{\tiny{RM-PC}}}(x,p)
\end{bmatrix}^T
\begin{bmatrix}
	0 & 1/2I\\
	1/2I & 0
\end{bmatrix}
\begin{bmatrix}
	u\\
	\mathcal{V}^{\text{\tiny{RM-PC}}}(x,p)
\end{bmatrix}
\end{align*}
Hence, setting 
\begin{equation}
\label{eq:sigmaRM-PC}
-\mathfrak{S}^{\text{\tiny{RM-PC}}}(x,p) =  \frac{1}{\bar{\tau}}\sum_{i=1}^n \lambda_i \mathcal{V}^{\text{\tiny{RM-PC}}}_i(x,p) \gamma_i(p)
\end{equation}
it follows that, in order to show RM-PC protocols satisfying $n=2$ or (\ref{cond:RM-PC}) are $\delta$-passive, we can prove under $n=2$ or (\ref{cond:RM-PC}) that $\mathcal{S}^{\text{\tiny{RM-PC}}}$ and $\mathfrak{S}^{\text{\tiny{RM-PC}}}$ are non-negative and satisfy (\ref{eq:PWiseDPass_Stat1}), (\ref{eq:PWiseDPass_Stat2}). From the non-negativity of $\bar{\tau}$, $\phi$, $x$ and $\lambda$ we see that $\mathcal{S}^{\text{\tiny{RM-PC}}}$ is non-negative. Moreover, plugging $\mathcal{S}^{\text{\tiny{RM-PC}}}$ to~\cite[Lemma 4]{Park2018Payoff-Dynamic-} it follows that  $\mathcal{S}^{\text{\tiny{RM-PC}}}$ satisfies (\ref{eq:PWiseDPass_Stat1}). Thus we are left with the analysis of non-negativity of $\mathfrak{S}^{\text{\tiny{RM-PC}}}$ and conditions under which $\mathfrak{S}^{\text{\tiny{RM-PC}}}$ satisfy (\ref{eq:PWiseDPass_Stat2}). 

Remainder of the proof is partitioned to 2 steps. Step \textbf{(i)} discusses non-negativity of $\mathfrak{S}^{\text{\tiny{RM-PC}}}$ and step \textbf{(ii)} examines the validity of (\ref{eq:PWiseDPass_Stat2}).

\textbf{\underline{Step i:}} In this step we discuss non-negativity of $\mathfrak{S}^{\text{\tiny{RM-PC}}}$. Under our choice of $\delta$-storage function, results that we get for $n=2$ and $n\geq 3$ differ and we split our analysis for these two cases.

\underline{$n=2$:} Under $n=2$ we will show that $\mathfrak{S}^{\text{\tiny{RM-PC}}}$ is non-negative for all $\lambda\succ 0$. For this instance we have
\begin{align}
-\mathfrak{S}^{\text{\tiny{RM-PC}}}(x,p)&=\frac{1}{\bar{\tau}}(\lambda_1(x_2\lambda_2\phi_1(p_1-p_2)\nonumber \\
&\quad -x_1\lambda_1\phi_2(p_2-p_1))\psi_2(p_2-p_1)\nonumber\\
&\quad +\lambda_2(x_1\lambda_1\phi_2(p_2-p_1)\nonumber\\
&\quad -x_2\lambda_2\phi_1(p_1-p_2))\psi_1(p_1-p_2))\label{sigma^(PC)_n=2}
\end{align}
Let us analyze the cases $p_1=p_2$, $p_1>p_2$ and $p_1<p_2$ separately. When $p_1=p_2$, \eqref{sigma^(PC)_n=2} becomes 0 by the sign preservation of $\phi$. If we assume $p_1>p_2$, then \eqref{sigma^(PC)_n=2} becomes $-\lambda_2^2x_2\phi_1(p_1-p_2)\psi_1(p_1-p_2)/\bar{\tau}$ which is non-positive for all $x\in\mathbb{X}$ and $\lambda\succ 0$. If $p_1<p_2$, then \eqref{sigma^(PC)_n=2} becomes $-\lambda_1^2x_1\phi_2(p_2-p_1)\psi_2(p_2-p_1)/\bar{\tau}$ which is again non-positive for all $x\in\mathbb{X}$ and $\lambda\succ 0$. 

\underline{$n\geq 3$:} Results that we have for the $n=2$ and $n\geq 3$ differ in the sense that, when $n\geq 3$ we show non-negativity of $\mathfrak{S}^{\text{\tiny{RM-PC}}}$ only for RM-PC protocols satisfying (\ref{cond:RM-PC}). Hence, in what follows we assume that (\ref{cond:RM-PC}) holds. Let us denote $u=\max_{i\in\{1,\dots,n\}}\{\lambda_i\}$ and $l=\min_{i\in\{1,\dots,n\}}\{\lambda_i\}$, so (\ref{cond:RM-PC}) can be written as $u/l<\bar{\lambda}_\phi$. Notice that
\begin{align*}
&-\mathfrak{S}^{\text{\tiny{RM-PC}}}(x,p)\\
&=\frac{1}{\bar{\tau}}\sum_{i=1}^n\gamma_i(p)\lambda_i\left(\sum_{j=1}^n x_j\lambda_j\phi_i(p_i-p_j)\right)\\
&\quad - \frac{1}{\bar{\tau}}\sum_{i=1}^n\gamma_i(p)\lambda_i\left(\sum_{j=1}^n x_i\lambda_i\phi_j(p_j-p_i)\right)\\
&= \frac{1}{\bar{\tau}}\sum_{i=1}^n\sum_{j=1}^n x_j\phi_i(p_i-p_j)\lambda_j(\lambda_i\gamma_i(p)-\lambda_j\gamma_j(p))\\
&= \frac{1}{\bar{\tau}}\begin{bmatrix}
	\sum_{i=1}^n\phi_i(p_i-p_1)\lambda_1(\lambda_i\gamma_i(p)-\lambda_1\gamma_1(p)) \\
	\vdots \\
	\sum_{i=1}^n\phi_i(p_i-p_n)\lambda_n(\lambda_i\gamma_i(p)-\lambda_n\gamma_n(p))
\end{bmatrix}^T x
\end{align*}
Since $x\succeq 0$ it follows that $-\mathfrak{S}^{\text{\tiny{RM-PC}}}(x,p)$ is non-positive for all $u\succeq\lambda\succeq l$, $x\in \mathbb{X}$ and $p\in\mathbb{R}^n$ if and only if for all $u\succeq\lambda\succeq l$, $p\in \mathbb{R}^n$ and $k\in \{1,\dots,n\}$,
\begin{align*}
\sum_{i=1}^n\phi_i(p_i-p_k)\lambda_k(\lambda_i\gamma_i(p)-\lambda_k\gamma_k(p) ) \leq 0
\end{align*}
which is equivalent to
\begin{align}
\label{ineq:RM-PClambdanotopt}
\sup_{k\in \{1,\dots,n\}, p\in\mathbb{R}^n, u\succeq\lambda\succeq l}&\Bigg\{\sum_{i\in \{1,\dots,n\}\setminus\{k\}}\phi_i(p_i-p_k) \nonumber \\
& \lambda_k(\lambda_i\gamma_i(p)-\lambda_k\gamma_k(p) )\Bigg\} \leq 0
\end{align}
We can take supremum with respect to one set of the variables, and then take the supremum of the resulting expression with respect to the ones left~\cite{boyd_vandenberghe_2004}. We first choose to take supremum with respect to $\lambda$. Fixing any $k\in\{1,\dots,n\}$ and $p\in\mathbb{R}^n$, since $\gamma_i$ and $\phi_i$ are non-negative for all $i\in \{1,\dots,n\}$, the expression on the left-hand side of (\ref{ineq:RM-PClambdanotopt}) is maximized with respect to $\lambda$ when $\lambda_i/
\lambda_k$ is maximized for all $i\in \{1,\dots,n\}\setminus\{k\}$. Due to the box constraint $u/l\geq\lambda_i/
\lambda_k\geq l/u$, we have that for any $i\in \{1,\dots,n\}$, supremum of $\lambda_i/
\lambda_k$ is reached when $\lambda_i/\lambda_k = u/l$. Thus (\ref{ineq:RM-PClambdanotopt}) holds if and only if the following holds:
\begin{multline}
\label{ineq:RM-PClambdaopt}
\sum_{i=1}^n \phi_i(p_i-p_k)\left(\frac{u}{l}\gamma_i(p)-\gamma_k(p) \right) \leq 0, \\ k\in\{1,\dots,n\}, \ p\in\mathbb{R}^n
\end{multline}
Notice that if $\sum_{i=1}^n\phi_i(p_i-p_k)\gamma_i(p)=0$, then (\ref{ineq:RM-PClambdaopt}) is satisfied, meaning (\ref{ineq:RM-PClambdaopt}) holds if and only if
\begin{align}
\frac{u}{l} \leq \inf\limits_{\substack{{k\in\{1,\dots,n\}, p\in\mathbb{R}^n}\\{ \sum_{i=1}^{n}\phi_i(p_i-p_k)\gamma_i(p)\neq 0}}}\left\{\frac{\gamma_k(p)\sum_{i=1}^{n}\phi_i(p_i-p_k)}{\sum_{i=1}^n\phi_i(p_i-p_k)\gamma_i(p)}\right\} \label{PC_delta_passivity_2}
\end{align}
On the account that (\ref{cond:RM-PC}) holds we get \eqref{PC_delta_passivity_2} is satisfied with strict inequality, in turn implying $\mathfrak{S}^{\text{\tiny{RM-PC}}}(x,p)\geq 0$ for all $x\in\mathbb{X}$ and $p\in\mathbb{R}^n$.	

\textbf{\underline{Step ii:}} In the second step we discuss under what conditions $\mathfrak{S}^{\text{\tiny{RM-PC}}}$ satisfies (\ref{eq:PWiseDPass_Stat2}). Similar to that of the conclusions on non-negativity of $\mathfrak{S}^{\text{\tiny{RM-PC}}}$, under our choice of $\delta$-storage function, results that we obtain for the $n=2$ and $n\geq 3$ cases differ. We divide our analysis for these two cases.

\underline{$n=2$:} Assuming $n=2$, we show that $\mathfrak{S}^{\text{\tiny{RM-PC}}}(x,p)=0$ if and only if $\mathcal{V}^{\text{\tiny{RM-PC}}}(x,p)=0$ for all $\lambda\succ 0$. We present a proof by analyzing the cases $p_1=p_2$, $p_1>p_2$ and $p_1<p_2$ separately. Recall that when $n=2$, $\mathfrak{S}^{\text{\tiny{RM-PC}}}$ is given by \eqref{sigma^(PC)_n=2}. If $p_1=p_2$, then \eqref{sigma^(PC)_n=2} is 0, but in this case $\mathcal{V}^{\text{\tiny{RM-PC}}}(x,p)=0$. Now assume $p_1>p_2$. Then, $\mathfrak{S}^{\text{\tiny{RM-PC}}}(x,p)$ becomes $-\lambda_2^2x_2\phi_1(p_1-p_2)\psi_1(p_1-p_2)/\bar{\tau}$, but since $\phi_1(p_1-p_2)>0$ and $\psi_1(p_1-p_2)>0$ we see that $\mathfrak{S}^{\text{\tiny{RM-PC}}}(x,p)=0$ implies $x_2=0$. Moreover, from $p_1>p_2$, we have $\phi_2(p_2-p_1)=0$. These combined yield $\mathcal{V}^{\text{\tiny{RM-PC}}}(x,p)=0$. For the case $p_2>p_1$, $\mathfrak{S}^{\text{\tiny{RM-PC}}}(x,p)$ becomes $-\lambda_1^2x_1\phi_2(p_2-p_1)\psi_2(p_2-p_1)/\bar{\tau}$. But since $\phi_2(p_2-p_1)>0$ and $\psi_2(p_2-p_1)>0$ we see that $\mathfrak{S}^{\text{\tiny{RM-PC}}}(x,p)=0$ implies $x_1=0$. From $p_2>p_1$, we also have $\phi_1(p_1-p_2)=0$. These combined again yield $\mathcal{V}^{\text{\tiny{RM-PC}}}(x,p)=0$. Hence, we arrive at $\mathfrak{S}^{\text{\tiny{RM-PC}}}(x,p)=0$ implies $\mathcal{V}^{\text{\tiny{RM-PC}}}(x,p)=0$. For the other direction, assume $\mathcal{V}^{\text{\tiny{RM-PC}}}(x,p)=0$. Then, since $\mathfrak{S}^{\text{\tiny{RM-PC}}}(x,p)=\sum_{i=1}^n\lambda_i \mathcal{V}^{\text{\tiny{RM-PC}}}_i(x,p)\gamma_i(p)$, it follows that $\mathfrak{S}^{\text{\tiny{RM-PC}}}(x,p)=0$. As a result $\mathfrak{S}^{\text{\tiny{RM-PC}}}(x,p)=0$ if and only if $\mathcal{V}^{\text{\tiny{RM-PC}}}(x,p)=0$.

\underline{$n\geq 3$:} Now assume $n\geq 3$. We will show that for all RM-PC protocols satisfying (\ref{cond:RM-PC}) we have $\mathfrak{S}^{\text{\tiny{RM-PC}}}(x,p) = 0$ if and only if $\mathcal{V}^{\text{\tiny{RM-PC}}}(x,p) =0$. Recall that
\begin{align*}
&\mathfrak{S}^{\text{\tiny{RM-PC}}}(x,p)=- \sum_{i=1}^n\sum_{j=1}^n \frac{x_j}{\bar{\tau}}\phi_i(p_i-p_j)\lambda_j(\lambda_i\gamma_i(p)-\lambda_j\gamma_j(p))
\end{align*}
Given any $j\in \{1,\dots,n\}$, there are three possibilities: out of $p_1,\dots,p_n$ it must be that, $p_j$ is the largest, $p_j$ is the second largest, or there exist $l,m\in\{1,\dots,n\}\setminus\{j\}$ such that $p_m>p_l>p_j$. We analyze these three cases separately. If $j$ is such that $p_j$ is the largest, then for all $i\in \{1,\dots,n\}$, $\phi_i(p_i-p_j)\lambda_j (\lambda_i\gamma_i(p)-\lambda_j\gamma_j(p) ) = 0$, and any $x_j$ gives $x_j\phi_i(p_i-p_j)\lambda_j (\lambda_i\gamma_i(p)-\lambda_j\gamma_j(p) )/\bar{\tau} =0$. In the second case, $p_j$ is the second largest. Let us denote $\mathcal{I}=\{i\in\{1,\dots,n\}:p_i>p_j\}$, so $\mathcal{I}$ is the set of strategies having greater payoff than that of $j$. For any $l\in\mathcal{I}$ we have that $\gamma_l(p)=\sum_{k=1}^n\psi_k(p_k-p_l)=0$ and $\gamma_j(p) = \sum_{k=1}^n \psi_k(p_k-p_j)\geq \psi_l(p_l-p_j) > 0$. However, for any $k\in \{1,\dots,n\}\setminus \mathcal{I}$ we have $\phi_k(p_k-p_j)=0$, implying $\phi_k(p_k-p_j)\lambda_k(\lambda_k\gamma_k(p)-\lambda_j\gamma_j(p))=0$. Therefore,
\begin{align*}
&\sum_{i=1}^n\phi_i(p_i-p_j)\lambda_j(\lambda_i\gamma_i(p)-\lambda_j\gamma_j(p))\frac{1}{\bar{\tau}}\\
&= \sum_{k\in \{1,\dots,n\}\setminus\mathcal{I}}\phi_k(p_k-p_j)\lambda_k(\lambda_k\gamma_k(p)-\lambda_j\gamma_j(p))\frac{1}{\bar{\tau}}\\
&+\quad\sum_{l\in \mathcal{I}}\phi_l(p_l-p_j)\lambda_l(\lambda_l\gamma_l(p)-\lambda_j\gamma_j(p))\frac{1}{\bar{\tau}}\\
&<0
\end{align*}
Finally, if $j$ is such that there exist $l,m\in\{1,\dots,n\}\setminus\{j\}$ with $p_m>p_l>p_j$, then $\gamma_l(p) = \sum_{k=1}^n \psi_k(p_k-p_l)>0$, thus $\sum_{i=1}^n\phi_i(p_i-p_j)\gamma_i(p)\geq \phi_l(p_l-p_j)\gamma_l(p) >0$. Consequently, (\ref{cond:RM-PC}) can be utilized to arrive at the following. For all $p\in\mathbb{R}^n$ such that there exists $l,m\in\{1,\dots,n\}\setminus\{j\}$ with $p_m>p_l>p_j$, it holds that $(u/l)<(\gamma_j(p)\sum_{i=1}^n\phi_i(p_i-p_j))/(\sum_{i=1}^n\phi_i(p_i-p_j)\gamma_i(p))$. This implies that for all $p\in\mathbb{R}^n$ with $l,m\in\{1,\dots,n\}\setminus\{j\}$ satisfying $p_m>p_l>p_j$ we have
\begin{align*}
&\sum_{i=1}^n\phi_i(p_i-p_j)\lambda_j(\lambda_i\gamma_i(p)-\lambda_j\gamma_j(p))\frac{1}{\bar{\tau}}\\
&\leq \sum_{i=1}^n\phi_i(p_i-p_j)l\left(\frac{u}{l}\gamma_j(p)-\gamma_i(p)\right)\frac{1}{\bar{\tau}}\\
&<0
\end{align*} 
From the analysis of these three cases on $p_j$, it becomes evident that $\mathfrak{S}^{\text{\tiny{RM-PC}}}(x,p)=\sum_{j=1}^nx_j\sum_{i=1}^n\phi_i(p_i-p_j)\lambda_j(\lambda_i\gamma_i(p)-\lambda_j\gamma_j(p))/\bar{\tau}=0$ if and only if $x_j>0$ only when $j\in \arg \max_{k\in \{1,\dots,n\}}\{p_k \}$. Hence, $\mathfrak{S}^{\text{\tiny{RM-PC}}}(x,p)=0$ if and only if $x\in \arg \max_{y\in\mathbb{X}}y^Tp$. Finally, by the Nash stationarity of RM-PC protocols we arrive at $\mathfrak{S}^{\text{\tiny{RM-PC}}}(x,p)$ if and only if $\mathcal{V}^{\text{\tiny{RM-PC}}}(x,p)=0$. $\blacksquare$

\subsubsection{Proof of Proposition \ref{prop:RM-PCBound>1}}
\label{pf:RM-PCBound>1}

We assume that $n^r\geq 3$ and drop the $r$ superscript for notational convenience. Under Assumption~\ref{assump:prop1}, we can substitute $\phi_i$ with $\bar{\phi}$ (also denote $\bar{\psi}(\tilde{p}) = \smallint_0^{\tilde{p}} \bar{\phi}(s)ds$ for $\tilde{p}\in\mathbb{R}$ and $\bar{\gamma}_i(p)=\sum_{k=1}^n\bar{\psi}(p_k-p_i)$ for $i\in\{1,\dots,n\}$, $p\in\mathbb{R}^n$) to rewrite \eqref{ineq:RM-PC} as
\begin{equation}\label{exp:RM-PC_expr}
\bar{\lambda}_{\phi}(n) = \min_{1 \leq k\leq n}~\inf_{p\in\Theta_k}  \mathcal{O}(k,p)
\end{equation}
where
\begin{align*}
&\mathcal{O}(k,p) := \frac{\bar{\gamma}_{k}(p)\sum_{i=1}^{n}\bar{\phi}(p_i-p_k)}{\sum_{i=1}^{n}\bar{\phi}(p_i-p_k)\bar{\gamma}_i(p)},\\
&\Theta_k := \left\{p\in\mathbb{R}^n \ \Big | \ \sum_{i=1}^n \bar{\phi}(p_i-p_k)\bar{\gamma}_i(p)\neq 0\right\}
\end{align*}
In what follows, we derive a lower bound to \eqref{exp:RM-PC_expr} that is greater than 1. Our approach consists of three steps. {\bf (i)}~First, we show that without loss of generality we can fix $k$ in \eqref{exp:RM-PC_expr} to be $n$, effectively discarding the minimization over $k$. Specifically, we will show that $\bar{\lambda}_{\phi}(n)=\inf_{p\in\Theta_n}\mathcal{O}(n,p)$. {\bf (ii)}~Then, we prove that the value of $\inf_{p\in\Theta_n}\mathcal{O}(n,p)$ is unchanged when we introduce the additional constraint $p_1\geq p_2\geq \dots \geq p_n$. {\bf (iii)}~Finally, by exploiting the fact that $\bar{\phi}$ is non-decreasing, we derive a lower bound to the value of $\inf_{p\in\Theta_n}\mathcal{O}(n,p)$ with the additional constraint $p_1\geq p_2\geq \dots \geq p_n$.

\noindent \textbf{\underline{Step i:}} We begin by showing that \eqref{exp:RM-PC_expr} is equal to $\inf_{p\in\Theta_n}\mathcal{O}(n,p)$. Fix any $k,l\in\{1,\dots,n\}$ and $p\in\Theta_k$. Let us construct $\tilde{p}$ by swapping the values of the $k$-th and $l$-th indices of $p$. Then, it follows that $\tilde{p}\in\Theta_l$ and $\mathcal{O}(k,p)=\mathcal{O}(l,\tilde{p})$. Therefore, infimum of $\mathcal{O}(k,p)$ over $p\in\Theta_k$ is independent of $k$, implying that without loss of generality we can fix the $k$ in \eqref{exp:RM-PC_expr} to be $n$ and discard the minimization over $k$. Hence, we can conclude from \eqref{exp:RM-PC_expr} that $\bar{\lambda}_{\phi}(n)=\inf_{p\in\Theta_n}\mathcal{O}(n,p)$.

\noindent \textbf{\underline{Step ii:}} Now we prove that the value of $\inf_{p\in\Theta_n}\mathcal{O}(n,p)$ does not change when the additional constraint $p_1\geq p_2\geq \dots \geq p_n$ is imposed on the problem. First, observe that for any given $p$ in $\Theta_n$, we have that $\mathcal{O}(n,p)=\mathcal{O}(n,\tilde{p})$ for any $\tilde{p}$ constructed from $p$ by arbitrarily permuting the first $n-1$ entries and leaving the $n$-th entry unchanged. Therefore, imposing the additional constraint $p_1\geq p_2\geq \dots\geq p_{n-1}$ would not change $\inf_{p\in\Theta_n}\mathcal{O}(n,p)$. Second, we will show that the infimum is unchanged even if we impose the more stringent constraint $p_1\geq p_2\geq\dots\geq p_{n-1}\geq p_n$. Specifically, we will show that given any $p$ in $\Theta_n$ with $p_1\geq \dots\geq p_{n-1}$, there exists a $\tilde{p}$ in $\Theta_n$ satisfying $\tilde{p}_1\geq\dots\geq\tilde{p}_{n-1}\geq\tilde{p}_n$ for which $\mathcal{O}(n,p)=\mathcal{O}(n,\tilde{p})$. To do so, take an arbitrary $p$ in $\Theta_n$ satisfying $p_1\geq \dots\geq p_{n-1}$. From $p\in\Theta_n$, it follows that $p_1\geq p_n$, since otherwise we would have $\sum_{i=1}^n\bar{\phi}(p_i-p_n)\bar{\gamma}_i(p)=0$. Thus, for any $p\in\Theta_n$ we either have $p_{n-1}\geq p_n$, or there is $m$ in $\{2,\dots,n-1\}$ such that $p_1\geq p_2\geq\dots\geq p_{m-1}\geq p_n> p_m\geq\dots\geq p_{n-1}$. If $p$ is such that $p_{n-1}\geq p_n$, then taking $\tilde{p}=p$ gives the desired result. On the other hand, if there is $m$ in $\{2,\dots,n-1\}$ such that $p_1\geq p_2\geq\dots\geq p_{m-1}\geq p_n> p_m\geq\dots\geq p_{n-1}$, then we construct $\tilde{p}$ by setting $\tilde{p}_i=p_n$ for all $i$ in $\{m,\dots,n\}$ and $\tilde{p}_j=p_j$ for all $j$ in $\{1,\dots,m-1\}$. One can now verify by direct substitution that for the constructed $\tilde{p}$ it holds that  $\mathcal{O}(n,p)=\mathcal{O}(n,\tilde{p})$ and $\tilde{p}$ is in $\Theta_n$, which concludes our proof of the second step.

Thus, defining the vectors $\tilde{\phi}_n(p)$ and $\bar{\gamma}(p)$ as
\begin{align*}
\tilde{\phi}_n(p) := \begin{bmatrix}
	(\tilde{\phi}_n)_1(p)\\
	\vdots\\
	(\tilde{\phi}_n)_n(p)\\
\end{bmatrix}= \begin{bmatrix}
	\frac{\bar{\phi}(p_1 - p_n)}{\sum_{i=1}^{n-1}\bar{\phi}(p_i-p_n) }\\
	\vdots\\
	\frac{\bar{\phi}(p_n - p_n)}{\sum_{i=1}^{n-1}\bar{\phi}(p_i-p_n) }\\
\end{bmatrix};  \bar{\gamma}(p) := \begin{bmatrix} \bar{\gamma}_1(p) \\ \vdots \\ \bar{\gamma}_n(p) \end{bmatrix}
\end{align*}
we have shown up to this point that 
\begin{equation}
\label{exp:RM-EPT_diss_equiv_simp}
\bar{\lambda}_{\phi}(n)=  \inf_{\substack{\tilde{\phi}_n^T(p)\bar{\gamma}(p)\neq 0 \\ p_1\geq \cdots \geq p_n }} \quad \frac{1}{\tilde{\phi}_n^T(p)\bar{\gamma}(p)/\bar{\gamma}_n(p)}
\end{equation}
Note that for $p\in\mathbb{R}^n$ satisfying $\tilde{\phi}_n^T(p)\bar{\gamma}(p)\neq 0$ and $p_1\geq \cdots \geq p_n$, there is $m\in\{1,\dots,n-1\}$ such that $p_m>p_n$, which in turn implies $\bar{\gamma}_n(p)=\sum_{k=1}^n\bar{\psi}(p_k-p_n)\geq \bar{\psi}(p_m-p_n)>0$.

\noindent \textbf{\underline{Step iii:}} As for the final step, we will derive a lower bound to \eqref{exp:RM-EPT_diss_equiv_simp} that is greater than 1. From the proof of~\cite[Theorem 7.2.9]{Sandholm2010Population-Game} it is known that for any $i,j\in \{1,\dots,n\}$, $p_i\geq p_j$ implies $\bar{\gamma}_i(p)\leq \bar{\gamma}_j(p)$, meaning under the constraint $p_1\geq \dots\geq p_n$ we have $0=\bar{\gamma}_1(p)\leq \dots\leq \bar{\gamma}_n(p)$. Thus, for all $p\in \mathbb{R}^n$ such that $p_1\geq \dots\geq p_n$ and $\sum_{i=1}^{n}\bar{\phi}(p_i-p_n)\bar{\gamma}_i(p)\neq 0$ we have
\begin{align}
\tilde{\phi}_n^T(p)\bar{\gamma}(p)/\bar{\gamma}_n(p) &= (\tilde{\phi}_n)_1(p)\cdot 0 + (\tilde{\phi}_n)_{2}(p)\frac{\bar{\gamma}_{2}(p)}{\bar{\gamma}_n(p)} \nonumber \\
&\quad + \dots + (\tilde{\phi}_n)_{n-1}(p)\frac{\bar{\gamma}_{n-1}(p)}{\bar{\gamma}_n(p)} + 0\cdot 1 \nonumber \\
&\leq 0+ (\tilde{\phi}_n)_{2}(p) + \dots + (\tilde{\phi}_n)_{n-1}(p) + 0 \nonumber \\
&= \sum_{i=2}^{n-1}(\tilde{\phi}_n)_{i}(p) \nonumber \\
&= 1-(\tilde{\phi}_n)_{1}(p)\label{ineq:RM-BNN_bound_simp}
\end{align}
The function $\bar{\phi}$ being non-decreasing implies under the constraints $p_1\geq \dots\geq p_n$ and $\sum_{i=1}^{n}\bar{\phi}(p_i-p_n)\bar{\gamma}_i(p)\neq 0$ that $(\tilde{\phi}_n)_{1}(p) \geq 1/(n-1)$. As a result 
\begin{align*}
\tilde{\phi}_n^T(p)\bar{\gamma}(p)/\bar{\gamma}_n(p) &\leq 1-\frac{1}{n-1}\\
&= \frac{n-2}{n-1}
\end{align*}
meaning $(n-1)/(n-2)$ is a lower bound to (\ref{exp:RM-EPT_diss_equiv_simp}).$\blacksquare$

\bibliographystyle{IEEEtran}
\bibliography{MartinsRefs,KaraRefs}

\begin{IEEEbiography}[{\includegraphics[width=1in,height=1.25in,clip,keepaspectratio]{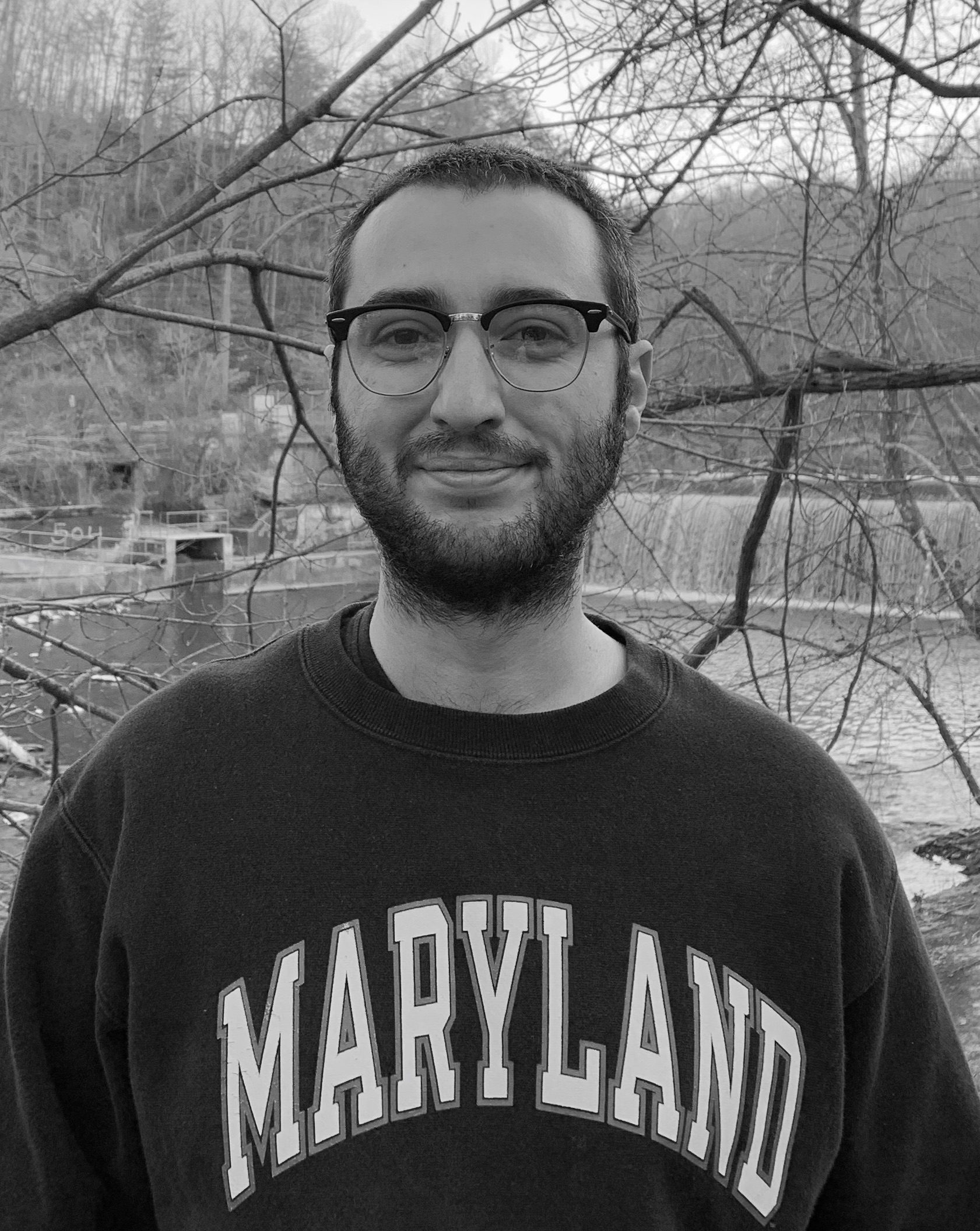}}]{Semih Kara} received his B.S. degree in Electrical and Electronics Engineering from Bilkent University, Turkey in 2017. He is currently pursuing a Ph.D. degree in Electrical and Computer Engineering from the University of Maryland, College Park.
\end{IEEEbiography}

\begin{IEEEbiography}[{\includegraphics[width=1in,height=1.25in,clip,keepaspectratio]{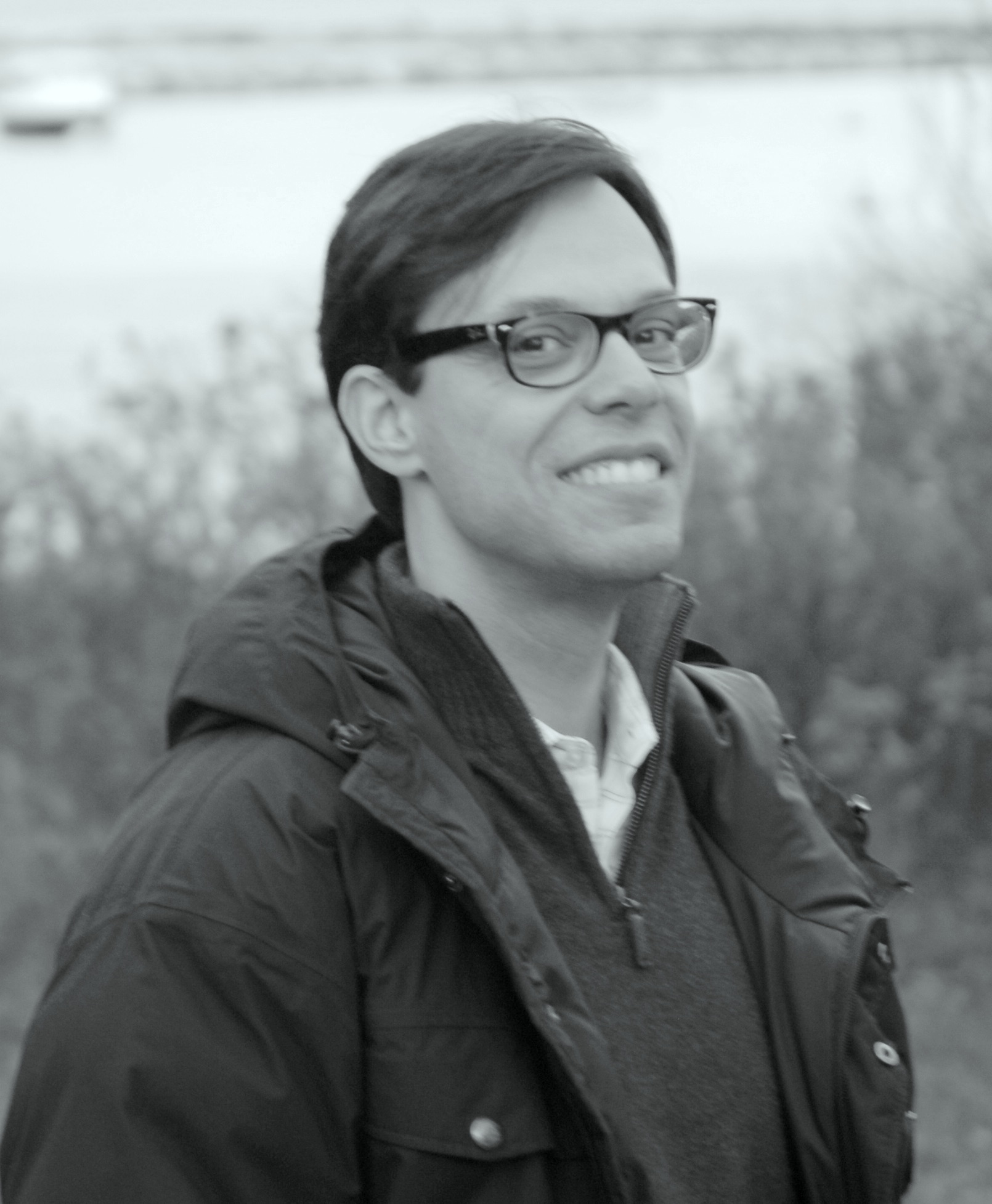}}]{Nuno Miguel Lara Cintra Martins} graduated with a M.S. degree in Electrical Engineering from I.such that, Portugal, in 1997, and a Ph.D. degree in Electrical Engineering and Computer Science with a minor in Mathematics from MIT in 2004. He has also concluded a Financial Technology Option program at Sloan School of Management (MIT) in 2004. He is Professor in the ECE Dept. of the U. of Maryland at College Park, where he also holds a joint appointment with the ISR. 
\end{IEEEbiography}

\end{document}